\documentclass[a4paper,11pt]{article}
\usepackage{jheppub} 
\usepackage{lineno}
\usepackage{orcidlink}

\title{\boldmath Study of $e^+e^- \to \pi^+\pi^-\Upsilon(1D)$ at Belle II}

\collaboration{The Belle II Collaboration}
\collaborationImg{\includegraphics[height=0.5in]{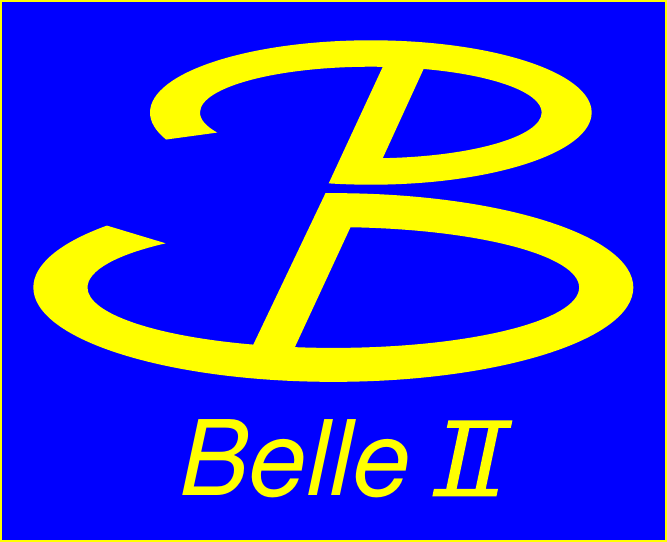}}
  \author{M.~Abumusabh\,\orcidlink{0009-0004-1031-5425},} 
  \author{I.~Adachi\,\orcidlink{0000-0003-2287-0173},} 
  \author{A.~Aggarwal\,\orcidlink{0000-0002-5623-3896},} 
  \author{L.~Aggarwal\,\orcidlink{0000-0002-0909-7537},} 
  \author{H.~Ahmed\,\orcidlink{0000-0003-3976-7498},} 
  \author{Y.~Ahn\,\orcidlink{0000-0001-6820-0576},} 
  \author{H.~Aihara\,\orcidlink{0000-0002-1907-5964},} 
  \author{S.~Alghamdi\,\orcidlink{0000-0001-7609-112X},} 
  \author{M.~Alhakami\,\orcidlink{0000-0002-2234-8628},} 
  \author{A.~Aloisio\,\orcidlink{0000-0002-3883-6693},} 
% ...

\abstract{
  The bottomonium spectrum, consisting of bound states of a $  b  $ quark and an anti-$  b  $ quark, provides an excellent laboratory for probing quantum chromodynamics in the non-perturbative regime. While S and P-wave bottomonium states are well studied experimentally, information on D-wave states remains scarce.
  We search for the D-wave bottomonium states $  \Upsilon_2(1D)  $ and $  \Upsilon_3(1D)  $ via the decay of a vector bottomonium-like state \(\Upsilon(10753)\) in the reaction $e^+e^- \to \pi^+\pi^- \Upsilon(1D)$, using $19.6~\mathrm{fb}^{-1}$ of data collected with the Belle II detector at center-of-mass energies $\sqrt{s} = 10.653$, 10.701, 10.745, and 10.805~GeV, in the vicinity of the $  \Upsilon(10753)  $ resonance. 
  No significant signals are observed.
  We set upper limits at the 90\% credibility level on the products of the cross sections and branching fractions, \(\sigma[e^+e^- \to \pi^+\pi^- \Upsilon_2(1D)] \times \mathcal{B}(\Upsilon_2(1D) \to \gamma \chi_{b1})\) and \(\sigma[e^+e^- \to \pi^+\pi^- \Upsilon_3(1D)] \times \mathcal{B}(\Upsilon_3(1D) \to \gamma \chi_{b2})\), at each center-of-mass energy.
}

\begin{document} 

\maketitle
\flushbottom

\section{Introduction}

Bottomonium states, composed of a bottom quark (\(b\)) and an antiquark (\(\bar{b}\)), offer a vital platform for probing quantum chromodynamics (QCD) in the non-perturbative regime~\cite{Brambilla:2010cs}.
The substantial mass of the bottom quark (\(\sim 4.18 \, \text{GeV}/c^2\)) leads to non-relativistic dynamics, positioning bottomonium as an ideal system for validating potential models, such as the Godfrey-Isgur relativized quark model~\cite{Godfrey:1985xj} and the Cornell potential model~\cite{Eichten:1978tg}.
The bottomonium spectrum includes S-wave states (with orbital angular momentum \(L=0\)), P-wave states (\(L=1\)), D-wave states (\(L=2\)) and higher orbital excitations. While S-wave and P-wave states have been extensively studied~\cite{pdg}, much less is known about D-wave states. Completing the bottomonium spectrum not only tests lattice QCD calculations but also rigorously evaluates our comprehension of bottomonium within the quark model framework~\cite{Godfrey:1985xj,Brambilla:2010cs,Patrignani:2012an,Eichten:2007qx}, potentially revealing unexpected states that could hint at exotic structures.

The \(1D\) states of the D-wave bottomonia, characterized by radial quantum number \(n=1\), include the spin-triplet configurations (\(1^3D_J\)) with total angular momentum \(J=1, 2, 3\) and the spin-singlet state (\(1^1D_2\)) with \(J=2\). 
The CLEO Collaboration reported the first observation for the $\Upsilon_J(1D)$ states (corresponding to the $1^3D_J$) via the radiative decay $\Upsilon(3S) \to \gamma \gamma \Upsilon_J(1D)$~\cite{CLEO:2004npj}, favoring a $J=2$ assignment among $J=1,2,3$. 
This observation was subsequently confirmed by the BaBar Collaboration through the same decay process~\cite{BaBar:2010tqb}, providing the first definitive identification of the $\Upsilon_2(1D)$ (\(1^3D_2\)) state.
Both experiments yielded inconclusive evidence for the \(\Upsilon_1(1D)\) (\(1^3D_1\)) and \(\Upsilon_3(1D)\) (\(1^3D_3\)) states. A theoretical prediction from the nonrelativistic screened-potential model estimates only small mass differences of \(M_{\Upsilon_3(1D)} - M_{\Upsilon_2(1D)} \approx 4 \, \text{MeV}\) and \(M_{\Upsilon_2(1D)} - M_{\Upsilon_1(1D)} \approx 7 \, \text{MeV}\)~\cite{yjd-BR}, implying that the observed signals may represent a composite of contributions from the \(\Upsilon_1(1D)\), \(\Upsilon_2(1D)\), and \(\Upsilon_3(1D)\) states. Additional experimental insights into the D-wave states have emerged from the Belle experiment, which saw a low-significance (\(2.4\sigma\)) excess of the decay \(\Upsilon(5S) \to \pi^+ \pi^- \Upsilon_J(1D)\) (\(J = 1, 2, 3\)) in an inclusive analysis of \(\Upsilon(5S) \to \pi^+ \pi^- X\) decays~\cite{y5s-pipiInclusive-belle}.
However, these analyses did not resolve the individual contributions of the \(\Upsilon_1(1D)\), \(\Upsilon_2(1D)\), and \(\Upsilon_3(1D)\) states, highlighting the necessity for further experimental efforts to elucidate their distinct properties.

In addition, exploring the \(\Upsilon_J(1D)\) states in decays from other resonances beyond \(\Upsilon(3S)\) and \(\Upsilon(5S)\) presents an intriguing avenue for discovery, potentially revealing new production mechanisms and transition dynamics in the bottomonium spectrum.
The \(\Upsilon(10753)\), a vector bottomonium-like state with quantum numbers (\(J^{PC} = 1^{--}\)), was first observed in \(e^+e^-\) collisions through the processes \(e^+e^- \to \pi^+\pi^- \Upsilon(nS)\) (\(n = 1, 2, 3\)) by the Belle Collaboration~\cite{Belle:pipiYnS}, with further confirmation and refined measurements by Belle II~\cite{Belle-II:pipiYnS,Belle-II:2025ubm,belle2-omegachibj,Belle-II:2025jus}. 
Additionally, its decay to $\omega \chi_{b1,b2}$, where $\chi_{b1}$ and $\chi_{b2}$ are P-wave bottomonium states, has been observed at Belle II~\cite{belle2-omegachibj}, suggesting that the $e^+e^- \to \omega \chi_{b1,b2}$ process observed near $\Upsilon(5S)$ from Belle~\cite{Belle:2014sys} could be due to the tail of the $\Upsilon(10753)$.
The production rate of $\omega \chi_{bJ}$ compared to $\pi^+\pi^- \Upsilon(nS)$ at the $\Upsilon(10753)$ significantly exceeds that at the $\Upsilon(5S)$, despite their identical quantum numbers ($J^{PC} = 1^{--}$) and small mass difference ($\sim 110$~MeV$/c^2$), which may indicate distinctly different internal structures for $\Upsilon(10753)$ and $\Upsilon(5S)$~\cite{belle2-omegachibj}. 
However, experimental data on the \(\Upsilon(10753)\) remain limited, leaving its nature unresolved.

There is a broad range of theoretical interpretations for the \(\Upsilon(10753)\), reflecting the complexity of its classification.
Conventional bottomonium models propose it to be a higher radial excitation~\cite{conventional-bottomonium1,conventional-bottomonium2,conventional-bottomonium3,conventional-bottomonium4,conventional-bottomonium5,conventional-bottomonium6,conventional-bottomonium7,conventional-bottomonium8,conventional-bottomonium9,conventional-bottomonium10}.
However, its measured mass~\cite{Belle-II:pipiYnS,Belle:pipiYnS,belle2-omegachibj} does not correspond to predicted masses of standard radial excitations~\cite{Godfrey:1985xj} (e.g., \(\Upsilon(5S)\) or \(\Upsilon(6S)\)). Furthermore, its decay rate to \(\omega \chi_{b1,b2}\)~\cite{belle2-omegachibj} is significantly enhanced compared to that of the \(\Upsilon(5S)\)~\cite{Belle:2014sys}.
Alternative hypotheses suggest it could be a hybrid state~\cite{hybrid1,hybrid2}, incorporating gluonic excitations, or a tetraquark~\cite{tetraquark1,tetraquark2,tetraquark3,tetraquark4}. These models predict distinct decay patterns and branching fractions, particularly for hadronic transitions, which could reveal signatures of exotic configurations.

The decay \(\Upsilon(10753) \to \pi^+\pi^- \Upsilon_J(1D)\) is kinematically feasible due to the mass difference of approximately 588 MeV, sufficient to produce two charged pions. In conventional bottomonium, dipion transitions are well-established, as seen in \(\Upsilon(nS) \to \pi^+\pi^- \Upsilon(mS)\) (\(n>m; n = 2, 3, 4; m = 1, 2\)) decays~\cite{pdg}. 
If \(\Upsilon(10753)\) is a conventional bottomonium state, potential models and non-relativistic QCD calculations predict that its dipion transition rate to the D-wave states \(\Upsilon_J(1D)\) would be suppressed by one to two orders of magnitude relative to the corresponding S-wave transitions, owing to the angular-momentum barrier and the reduced overlap of the radial wave functions~\cite{Godfrey:1985xj,Brambilla:2010cs,Patrignani:2012an,Eichten:2007qx}.
Nevertheless, the mechanism involving intermediate \(Z_b\) states enhances the dipion transition rate to \(\Upsilon_J(1D)\), yielding branching fractions of order \(10^{-3}\), comparable to those for S-wave dipion transitions~\cite{Wu:2020edh}.
Conversely, exotic interpretations for \(\Upsilon(10753)\) may yield different rates, or distinct dipion mass spectra, due to different wave functions or selection rules~\cite{tetraquark1}. The study of \(\Upsilon(10753) \to \pi^+\pi^- \Upsilon_J(1D)\) decay is critical for testing these models and elucidating the \(\Upsilon(10753)\)'s nature.

In this analysis, we search for the processes \(e^+e^- \to \Upsilon(10753) \to \pi^+\pi^- \Upsilon_J(1D)\) (\(J = 2, 3\)) at Belle II. The \(\Upsilon_J(1D)\) candidates are reconstructed via the decays \(\Upsilon_J(1D) \to \gamma \chi_{bJ'}\) (\(J' = 0, 1, 2\)), with \(\chi_{bJ'} \to \gamma \Upsilon(1S)\) and \(\Upsilon(1S) \to \ell^+\ell^-\) (\(\ell = e, \mu\)). 
By exploring these decay channels, we aim to probe the properties of the \(\Upsilon_J(1D)\) states and provide new insights into the internal structure of the \(\Upsilon(10753)\) resonance, advancing our understanding of bottomonium spectroscopy and the potential existence of exotic hadrons beyond bottomonium. 
The \(J = 1\) case, corresponding to the \(\Upsilon_1(1D)\) state, is excluded from this study. Theoretical predictions indicate that its decay favors \(\gamma \chi_{b0}\)~\cite{yjd-BR}, but the branching fraction for \(\chi_{b0}\to\gamma\Upsilon(1S)\) is exceedingly small~\cite{pdg} which results in limited sensitivity for this channel. For \(\Upsilon_2(1D)\) (\(J = 2\)), theoretical predictions suggest that it primarily decays to \(\gamma\chi_{b1}\) or \(\gamma\chi_{b2}\), with the branching fraction for \(\Upsilon_2(1D)\to\gamma\chi_{b1}\to\gamma\gamma\Upsilon(1S)\) being approximately six times greater than that for \(\Upsilon_2(1D)\to\gamma\chi_{b2}\to\gamma\gamma\Upsilon(1S)\)~\cite{yjd-BR}. To maximize sensitivity by focusing on the mode with the highest expected signal yield relative to background, we exclusively consider the \(\chi_{b1}\) decay for the \(\Upsilon_2(1D)\) search. In the case of \(\Upsilon_3(1D)\) (\(J = 3\)), the branching fraction for \(\Upsilon_3(1D)\to\gamma\chi_{b2}\) approaches 100\%~\cite{yjd-BR}, leading to a focus solely on this \(\chi_{b2}\) decay channel.
We measure the products of the cross sections and branching fractions for \(\sigma[e^+e^- \to \pi^+\pi^- \Upsilon_2(1D)] \times \mathcal{B}(\Upsilon_2(1D) \to \gamma \chi_{b1})\) and \(\sigma[e^+e^- \to \pi^+\pi^- \Upsilon_3(1D)] \times \mathcal{B}(\Upsilon_3(1D) \to \gamma \chi_{b2})\).

\section{Belle II Detector and Data Samples}
\label{sec:detector-data-samples}

The Belle~II experiment is located at SuperKEKB,
which collides electrons and positrons at and near the $\Upsilon(4S)$ resonance~\cite{Akai:2018mbz}. The Belle II
detector~\cite{Abe:2010gxa} has a cylindrical geometry and includes a two-layer silicon-pixel detector~(PXD) surrounded by a four-layer double-sided silicon-strip detector~(SVD)~\cite{Belle-IISVD:2022upf} and a 56-layer central drift chamber~(CDC). 
Position information from these detectors is used to reconstruct the trajectories of charged particles~\cite{Bertacchi:2020eez}.
Only one sixth of the second layer of the PXD was installed for the data analyzed here. 
The symmetry axis of these detectors, defined as the $z$-axis, is almost coincident with the direction of the electron beam.
Surrounding the CDC, which also provides $dE/dx$ energy-loss measurements, is a time-of-propagation counter~(TOP)~\cite{Kotchetkov:2018qzw} in the central region and an aerogel-based ring-imaging Cherenkov counter~(ARICH) in the forward region.  These detectors provide charged-particle identification. 
The TOP and ARICH are surrounded by an electromagnetic calorimeter (ECL) consisting of CsI(Tl) crystals that covers the central, forward and backward regions, primarily providing energy and timing measurements for photons and electrons.
Outside of the ECL is a superconducting solenoid magnet. 
The magnet provides a 1.5~T magnetic field that is parallel to the $z$-axis. 
Its flux return is instrumented with resistive-plate chambers and plastic scintillator modules to detect muons, $K^0_L$ mesons, and neutrons.

This analysis is performed using the Belle II Analysis Software Framework~\cite{Kuhr:2018lps,basf2-zenodo}.
The data sample corresponds to an integrated luminosity of \(19.6~\mathrm{fb}^{-1}\), recorded in November 2021 at center-of-mass energies near \(10.75~\mathrm{GeV}\) (\(\sqrt{s} = 10.653\), \(10.701\), \(10.745\), and \(10.805~\mathrm{GeV}\)).
These energies were chosen to probe the $\Upsilon(10753)$ resonance.

The Monte Carlo (MC) simulation samples are generated using a \textsc{geant4}-based software package~\cite{GEANT4:2002zbu}, which incorporates a detailed geometric description of the Belle II detector and its response. These simulations are essential for evaluating detection efficiencies, optimizing event selection criteria, and estimating background contributions.

At each center-of-mass energy, MC events for the signal processes \( e^+ e^- \to \pi^+ \pi^- \Upsilon_J(1D) \) (\(J = 2, 3\)), followed by \( \Upsilon_J(1D) \to \gamma \chi_{b1} \) or \( \gamma \chi_{b2} \), are produced using the \textsc{evtgen} package~\cite{Lange:2001uf}.
The dynamics of the $\pi^+\pi^- \Upsilon_J(1D)$ production and the $  \Upsilon_J(1D)  $ decay are modeled using a phase-space (PHSP) approach.
Initial-state radiation (ISR) is simulated at next-to-leading order accuracy in quantum electrodynamics using \textsc{phokhara}~\cite{Phokhara}, with the initial input cross section line-shape proportional to \( \frac{1}{s} \). The maximum energy of ISR photons is set by the production threshold of the \( \pi^+ \pi^- \Upsilon_J(1D) \) system. Final-state radiation from charged particles is incorporated using \textsc{photos}~\cite{Barberio:1990ms}.

To investigate specific background sources that yield similar or identical final state particles, MC samples are generated for processes including \( e^+ e^- \to \omega [\to \pi^+ \pi^- \pi^0] \chi_{b1,b2} [\to \gamma \Upsilon(1S)]\), \( e^+ e^- \to \gamma_{\text{ISR}} \Upsilon(2S) [\to \pi^+ \pi^- \Upsilon(1S)] \), and \( e^+ e^- \to \eta [\to \gamma \gamma] \Upsilon(2S) [\to \pi^+ \pi^- \Upsilon(1S)]\). Additional simulated samples of continuum \( e^+ e^- \to q \bar{q} \) (\( q = u, d, s, c \)) processes, low-multiplicity quantum electrodynamic processes (e.g., Bhabha scattering, \( \mu^+ \mu^- (\gamma) \), \( \tau^+ \tau^- (\gamma) \), and \(\gamma\gamma\))~\cite{Balossini:2006wc,Balossini:2008xr}, ISR-produced hadron pair processes~\cite{Phokhara}, and four-track processes with at least one lepton pair~\cite{Berends:1984gf,Berends:1986ig} are examined to assess possible background contributions.
Backgrounds from $B\bar{B}$ production are negligible at the center-of-mass energies studied.

\section{Event Selection and Background Study}

Events are selected online using a hardware trigger that leverages information from the CDC and the ECL~\cite{Iwasaki:2011za}. In the offline analysis, photon candidates are identified from clusters in the ECL within the angular acceptance of the CDC, spanning the polar angle range \(17^{\circ} < \theta < 150^{\circ}\). Each cluster must satisfy a quality condition where the sum of weights \(w_i\) (\(w_i \leq 1\), where \(w_i\) refers to the energy-weighted contribution from the \(i\)-th crystal associated with this cluster) of all contributing crystals exceeds 1.5. 
The sum equals the number of crystals in non-overlapping clusters but can be non-integer for energy-shared clusters.
This requirement is equivalent to demanding that the cluster receives contributions from more than one crystal.
The deposited energy in the cluster must exceed region-specific thresholds: \(20~\text{MeV}\) in the barrel region, \(22.5~\text{MeV}\) in the forward endcap region, or \(20~\text{MeV}\) in the backward endcap region of the ECL.
In the signal process, typical photon energies in the laboratory frame range from 100 to 600~MeV, resulting in no efficiency loss due to these thresholds, as confirmed by MC simulations.

Charged particles are identified using offline selection criteria applied to variables such as impact parameter, momentum, and ECL cluster energy. To ensure a track originates from the \(e^+e^-\) interaction point, we impose constraints: \(\sqrt{d_x^2 + d_y^2} < 2.0~\text{cm}\) and \(|d_z| < 4.0~\text{cm}\), where \(d_x\), \(d_y\), and \(d_z\) are the coordinates in the \(x\), \(y\), and \(z\) directions, respectively, for the point (on the track) with the closest approach to the interaction point (determined using di-muon events).
Pion (\(\pi^{\pm}\)) and lepton (\(\ell^{\pm}\)) candidates are distinguished by their momenta: \(p(\pi^{\pm}) < 1.0~\text{GeV}\) and \(p(\ell^{\pm}) > 1.0~\text{GeV}\), consistent with the kinematic properties of the signal process as verified by MC simulations.
Lepton identification exploits the fact that electrons typically deposit most of their energy in the ECL, whereas muons deposit little.
The \(e^{\pm}\) (\(\mu^{\pm}\)) candidates are therefore differentiated based on ECL cluster energy: \(E_{\text{cluster}}(e^{\pm}) > 1.0~\text{GeV}\) and \(E_{\text{cluster}}(\mu^{\pm}) < 1.0~\text{GeV}\).
In the signal processes, MC simulations indicate that the event-level efficiencies are 100.0\% for identifying $\pi^+\pi^-$, 97.4\% for the $e^+e^-$, and 96.7\% for the $\mu^+\mu^-$. Misidentification rates are minimal, with 0.1\% for $e^+e^-$ as $\mu^+\mu^-$, and 0.3\% for $\mu^+\mu^-$ as $e^+e^-$, and pions are never misidentified as leptons (or vice versa).

To account for energy losses from bremsstrahlung in relativistic electrons (positrons), photons (satisfying the selection criteria described above) within a \(0.05~\text{rad}\) cone around the electron's (positron's) initial momentum direction are identified, and their four-momenta are added to that of the electron (positron).
This correction is applied to electrons and positrons only.

Signal candidates are reconstructed from a pair of oppositely charged pions, a lepton pair (either \(e^+e^-\) or \(\mu^+\mu^-\)), and two photons.
To improve the signal resolution and suppress backgrounds with incompatible final-states, a kinematic fit~\cite{OrcaKinFit} is performed. The fit enforces four-momentum conservation by constraining the total four-momentum of the final state particles to equal that of the initial $e^+e^-$ collision.
Events with poor fit quality are rejected by requiring \(\chi^2/N_{\text{DOF}} < 25\), where \(\chi^2\) is the goodness-of-fit statistic of the kinematic fit, and \(N_{\text{DOF}}=4\) is the number of degrees of freedom in the fit.
This loose requirement is chosen to maximize signal efficiency while effectively removing events with obviously incompatible kinematics; per MC simulations, 97.3\% of signal events are retained.
After this requirement, the average number of candidates per event is 1.1, with 94.5\% of events yielding a single candidate. For the small fraction of events with multiple candidates, the candidate with the smallest \(\chi^2/N_{\text{DOF}}\) is selected for further analysis.

The \(\Upsilon(1S)\) state is reconstructed from the lepton pair in each signal candidate. MC simulations show similar \(M(\ell^+\ell^-)\) invariant mass distributions for \(e^+e^-\) and \(\mu^+\mu^-\) after the kinematic fit, allowing a unified selection window of \([9.430, 9.499]~\mathrm{GeV}/c^2\) for both modes. This window encompasses \(\pm 3\sigma\) around the \(\Upsilon(1S)\) mass, with \(\sigma\) denoting the mass resolution derived from MC simulations.

The \(\chi_{b1}\) and \(\chi_{b2}\) states are reconstructed by combining the \(\Upsilon(1S)\) with one of the two signal photons. 
The energy distributions of these two photons overlap significantly in the laboratory frame, complicating the assignment of each photon to its originating decay.
However, when evaluated in the \(\Upsilon_J(1D)\) rest frame, the photon energies become nearly monoenergetic, greatly reducing the Doppler broadening present in the laboratory frame. This enables a clear distinction between the lower-energy photon (\(\gamma_L\)) from the \(\Upsilon_J(1D) \to \gamma_L \chi_{b1}/\chi_{b2}\) decay, and the higher-energy photon (\(\gamma_H\)) from the \(\chi_{b1}/\chi_{b2} \to \gamma_H \Upsilon(1S)\) decay. 
Simulations indicate that mis-reconstruction due to photon swapping occurs in less than 0.1\% of cases, making its impact negligible.

The invariant mass \(M[\gamma_H \Upsilon(1S)]\) of the \(\chi_{b1}/\chi_{b2}\) candidate is defined as:
\begin{equation}
M[\gamma_H \Upsilon(1S)] \equiv M(\gamma_H \ell^+ \ell^-) - M(\ell^+ \ell^-) + m[\Upsilon(1S)],
\end{equation}
where \(m[\Upsilon(1S)]\) is the nominal mass~\cite{pdg}. This subtracts the lepton pair mass and adds the known \(\Upsilon(1S)\) mass, which enhances the resolution of the \(\chi_{b1}/\chi_{b2}\) mass peaks (7.9~\text{MeV}$/c^2$). Signal windows are defined as \([9.869, 9.915]~\mathrm{GeV}/c^2\) for \(\chi_{b1}\) and \([9.888, 9.934]~\mathrm{GeV}/c^2\) for \(\chi_{b2}\), corresponding to \(\pm 3\sigma\) around their respective masses.

Analysis of simulated background samples (as described in Section~\ref{sec:detector-data-samples}) reveals a significant accumulation of events near \(\cos(\theta_{\pi^+\pi^-}) = 1\), where \(\theta_{\pi^+\pi^-}\) is the opening angle between the two pions in the laboratory frame. These events arise from radiative Bhabha (\(e^+e^- \to e^+e^-\gamma\)) and di-muon (\(e^+e^- \to \mu^+\mu^-\gamma\)) processes in which the emitted photon converts to an \(e^+e^-\) pair that have a small opening angle. When these converted \(e^+e^-\) pairs are misidentified as \(\pi^+\pi^-\), they lead to the observed accumulation. To mitigate this background, we apply a requirement of \(\cos(\theta_{\pi^+\pi^-}) < 0.95\).
This cut eliminates essentially 100\% of these background events while retaining 97\% of the signal efficiency, as validated by MC simulations.

The process \(e^+e^- \to \eta \Upsilon(2S)\), with a significant cross section near \(\sqrt{s} = 10.75~\text{GeV}\)~\cite{Belle-II:2025ubm}, matches the final state studied in this analysis. MC simulations indicate that it produces a background peak in the key signal observable (the recoil mass against the $\pi^+\pi^-$ pair) near the \(\Upsilon_J(1D)\) signal region, necessitating its careful removal. We apply a veto on the photon pair invariant mass, excluding events where \(M(\gamma\gamma) \in [0.511, 0.576]~\text{GeV}/c^2\), corresponding to the \(\eta\) mass window. This criterion eliminates 99\% of the $\eta \Upsilon(2S)$ background while preserving 85\% of the signal efficiency, as determined from MC studies.

The primary remaining background after the event selection arises from \(e^+e^- \to \omega \chi_{b1}/\chi_{b2}\), a well-studied process at Belle~II~\cite{belle2-omegachibj}. This background, characterized by a broad distribution in the signal observable ($\pi^+\pi^-$ recoil mass), is modeled using MC simulations and incorporated into the fit as a fixed shape, as detailed in the subsequent section. Contributions from other background sources are found to be negligible.

\section{Signal Extraction}

To analyze the \(\Upsilon_J(1D)\) spectrum, we have evaluated two methods: the invariant mass of the \(\gamma\gamma\Upsilon(1S)\) system (after the kinematic fit) and the recoil mass against the \(\pi^+\pi^-\) pair (prior to the kinematic fit). 
The recoil mass is defined as \(M(\pi^+\pi^-)^{\rm recoil} \equiv \sqrt{(P_{e^+e^-} - P_{\pi^+} - P_{\pi^-})^2}\), where \(P_{e^+e^-} = (\sqrt{s},0,0,0)\) is the four-momentum of the initial \(e^+e^-\) system in the center-of-mass frame, and \(P_{\pi^+}\) and \(P_{\pi^-}\) are the four-momenta of the \(\pi^+\) and \(\pi^-\) candidates, respectively.
MC simulations reveal that the \(\pi^+\pi^-\) recoil mass spectrum offers superior resolution (5.9 MeV/\(c^2\)) compared to the \(\gamma\gamma\Upsilon(1S)\) invariant mass spectrum (6.1 MeV/\(c^2\)). 
This improvement stems from the Belle II detector's precise tracking capabilities for low-momentum charged pions (\(\pi^+\) and \(\pi^-\)). Consequently, we adopt the \(\pi^+\pi^-\) recoil mass spectrum for signal extraction to maximize sensitivity.

\begin{figure}[h]
  \centering
  \includegraphics[height=2.5in]{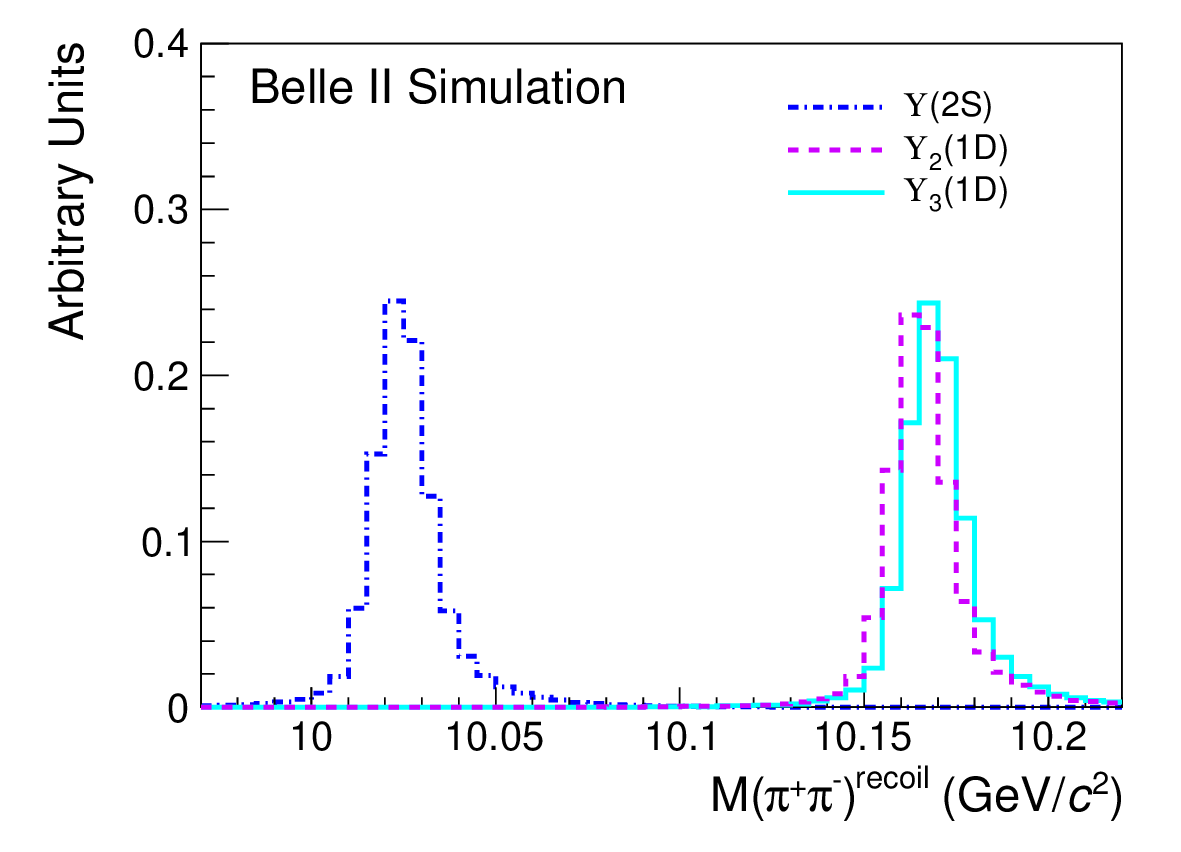}
  \caption{
    Normalized \(M(\pi^+\pi^-)^{\text{recoil}}\) distributions for signal MC events. 
    The dot-dashed blue line represents the \(\Upsilon(2S)\) signal, the dashed violet line is the \(\Upsilon_2(1D)\) signal, and the solid cyan line is the \(\Upsilon_3(1D)\) signal. The simulated peaks for the \(\Upsilon_2(1D)\) and \(\Upsilon_3(1D)\) states are separated by approximately 4~MeV/\(c^2\), consistent with theoretical predictions~\cite{yjd-BR,Segovia:2016xqb,Wang:2018rjg,Godfrey:2015dia}.
  }
  \label{pipiRecoilMC}
\end{figure}

\begin{figure}[h]
  \centering
  \includegraphics[height=2.1in]{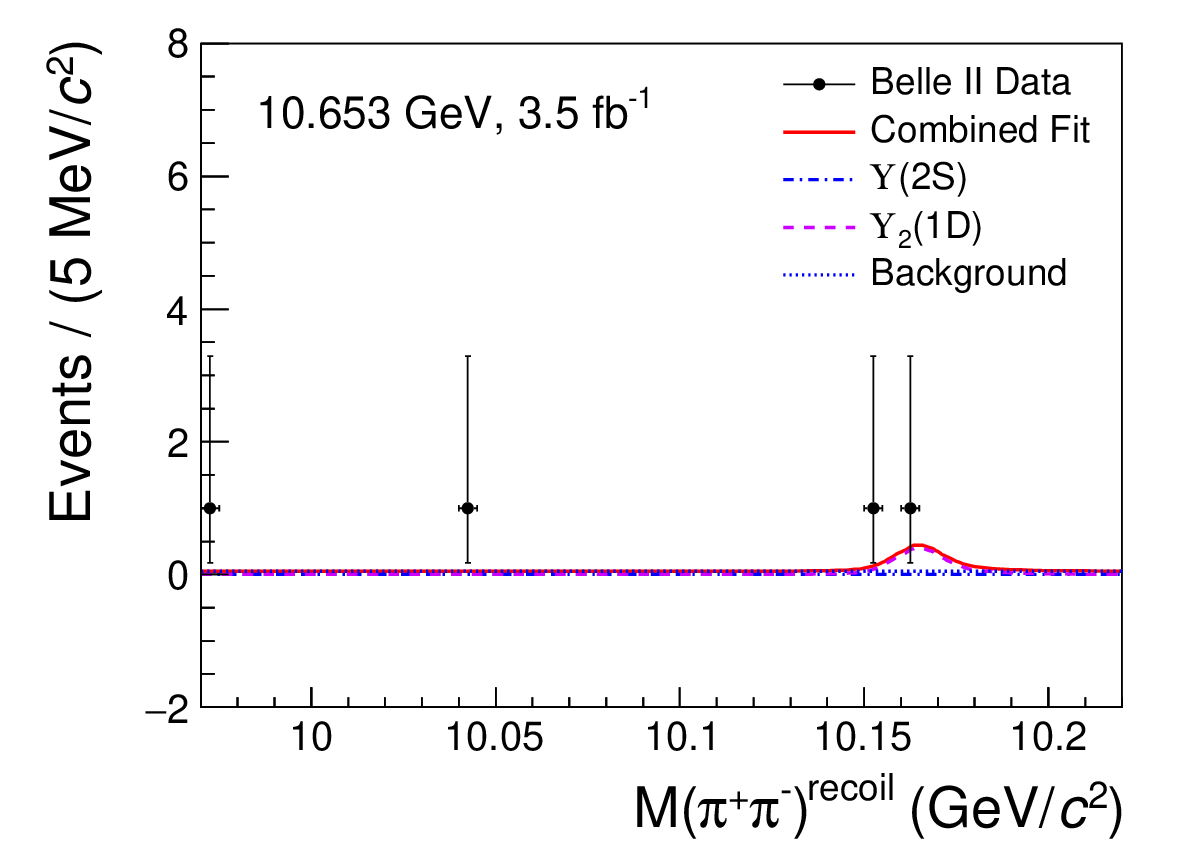}
  \includegraphics[height=2.1in]{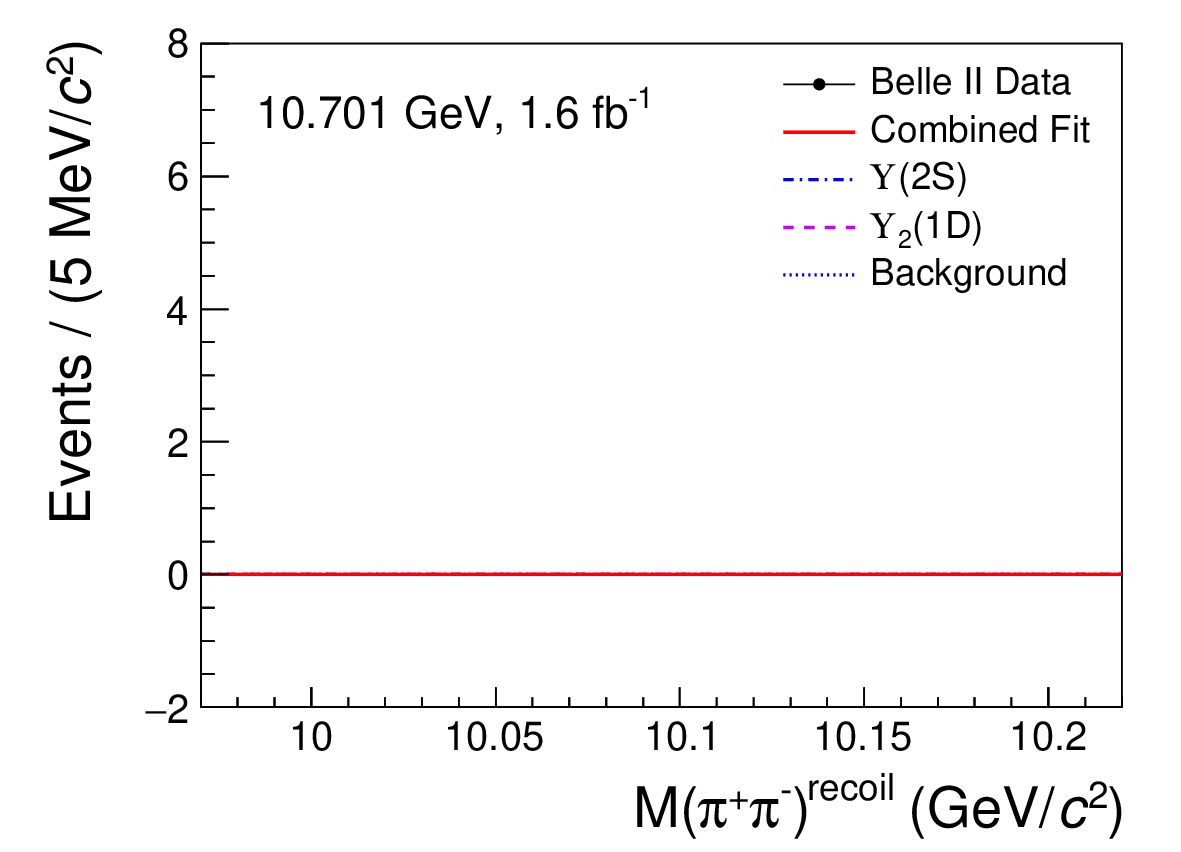}
  \includegraphics[height=2.1in]{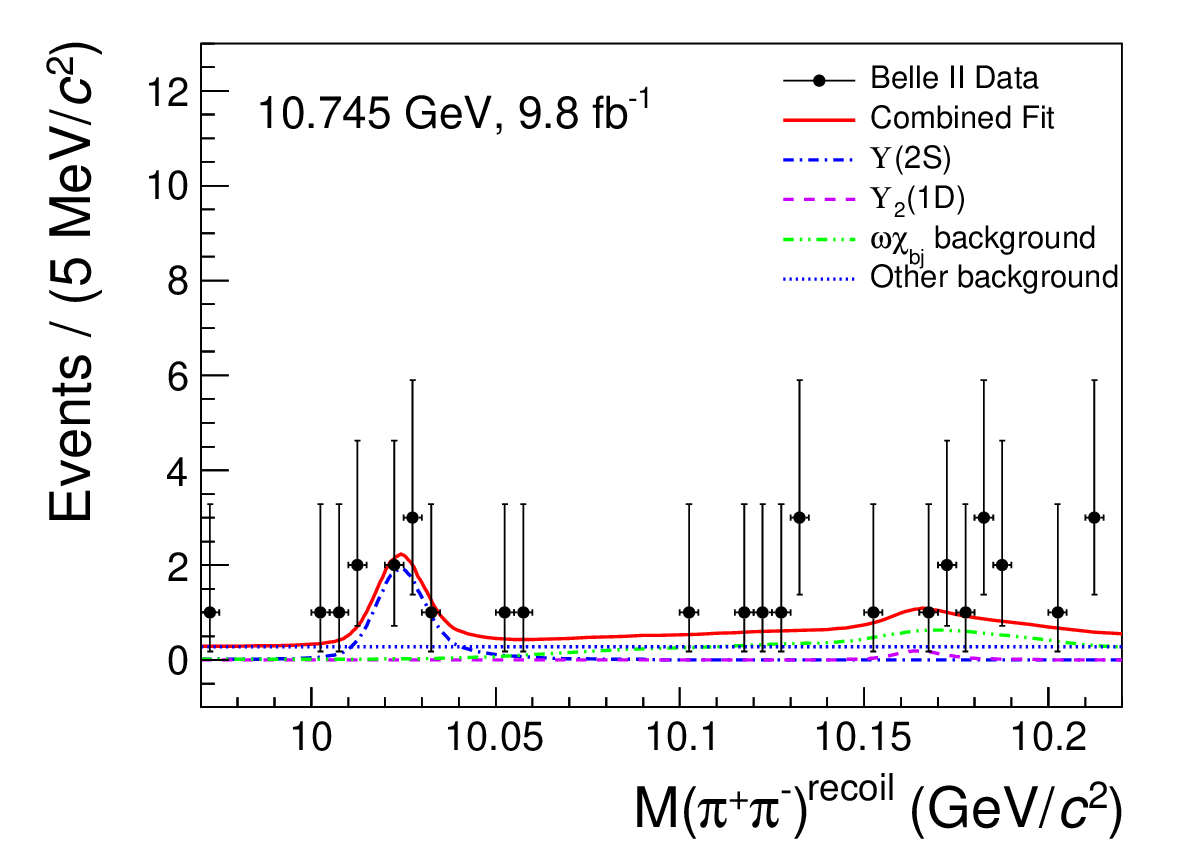}
  \includegraphics[height=2.1in]{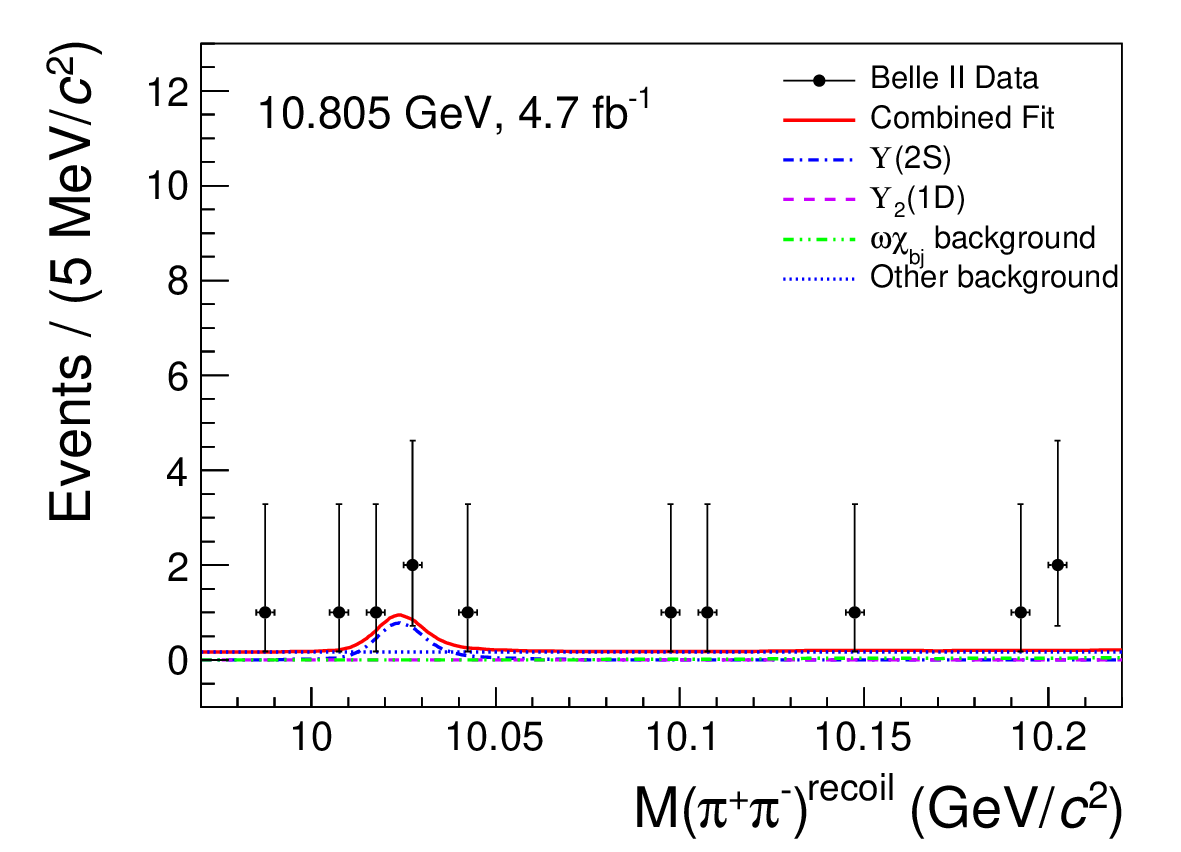}
  \caption{Fits to the \( M(\pi^+\pi^-)^{\text{recoil}} \) distributions for data events in the \( \chi_{b1} \) channel. 
  Dots with error bars represent the data, the solid red line is the total fit, the dashed violet line is the \( \Upsilon_2(1D) \) signal component, the dot-dashed blue line is the \( \Upsilon(2S) \) signal component, the double-dot-dashed green line represents the \( \omega \chi_{b1,b2} \) background, and the dotted blue line is the constant background component.
  Note that the \(\sqrt{s} = 10.701~\mathrm{GeV}\) sample contains zero events.
  }
  \label{pipiRecoilData}
\end{figure}

\begin{figure}[h]
  \centering
  \includegraphics[height=2.1in]{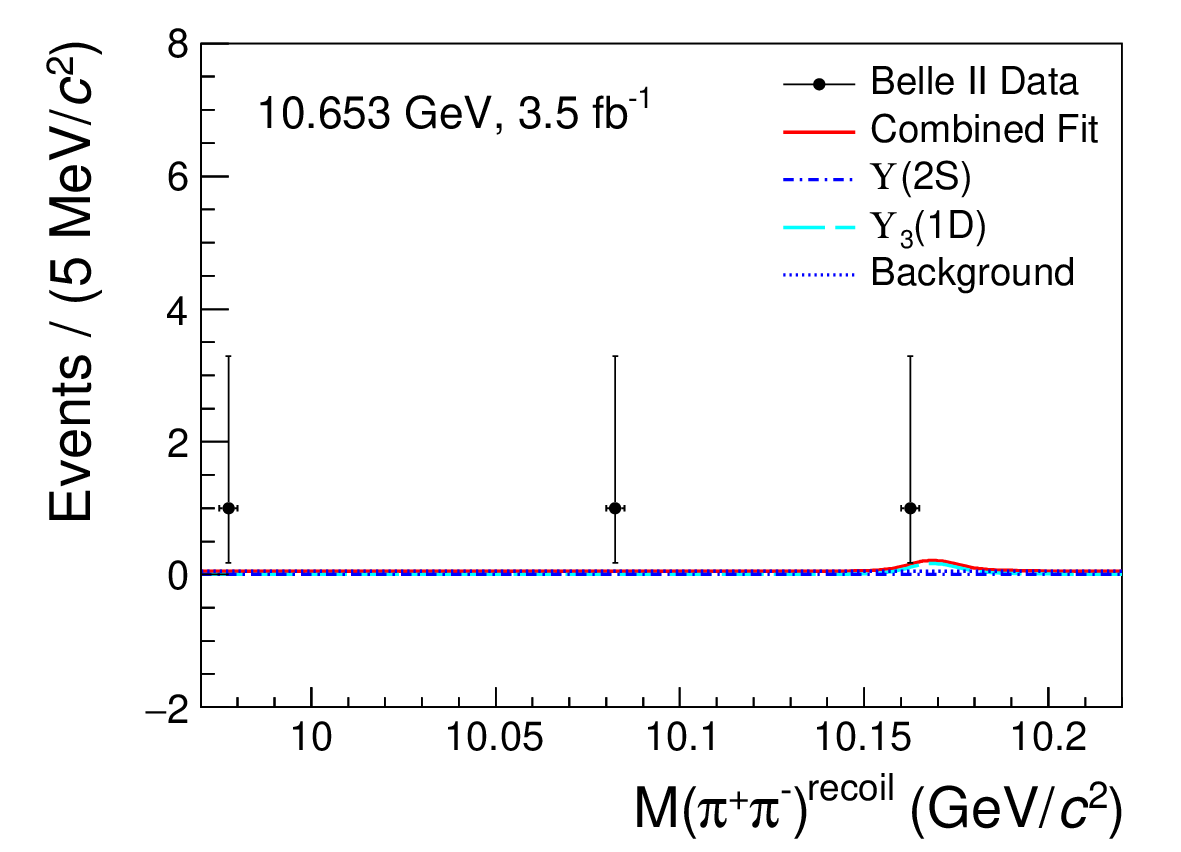}
  \includegraphics[height=2.1in]{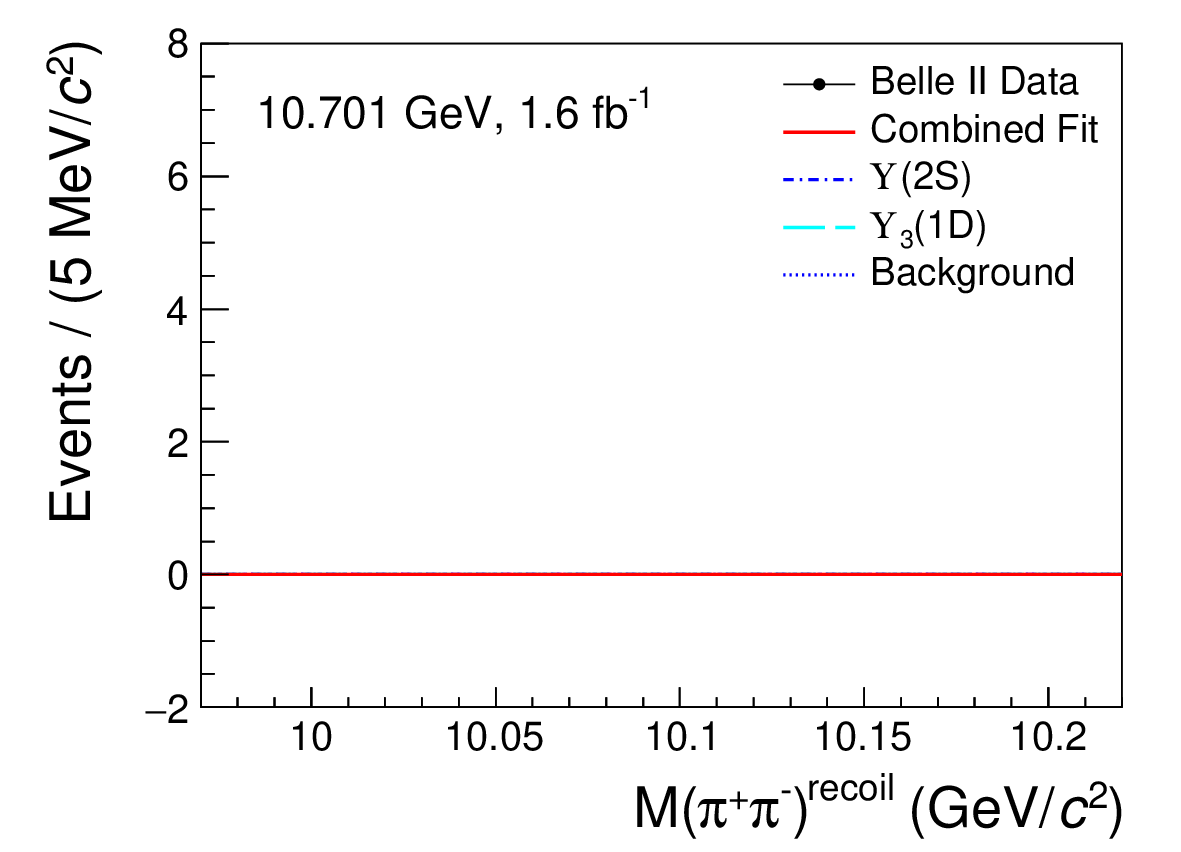}
  \includegraphics[height=2.1in]{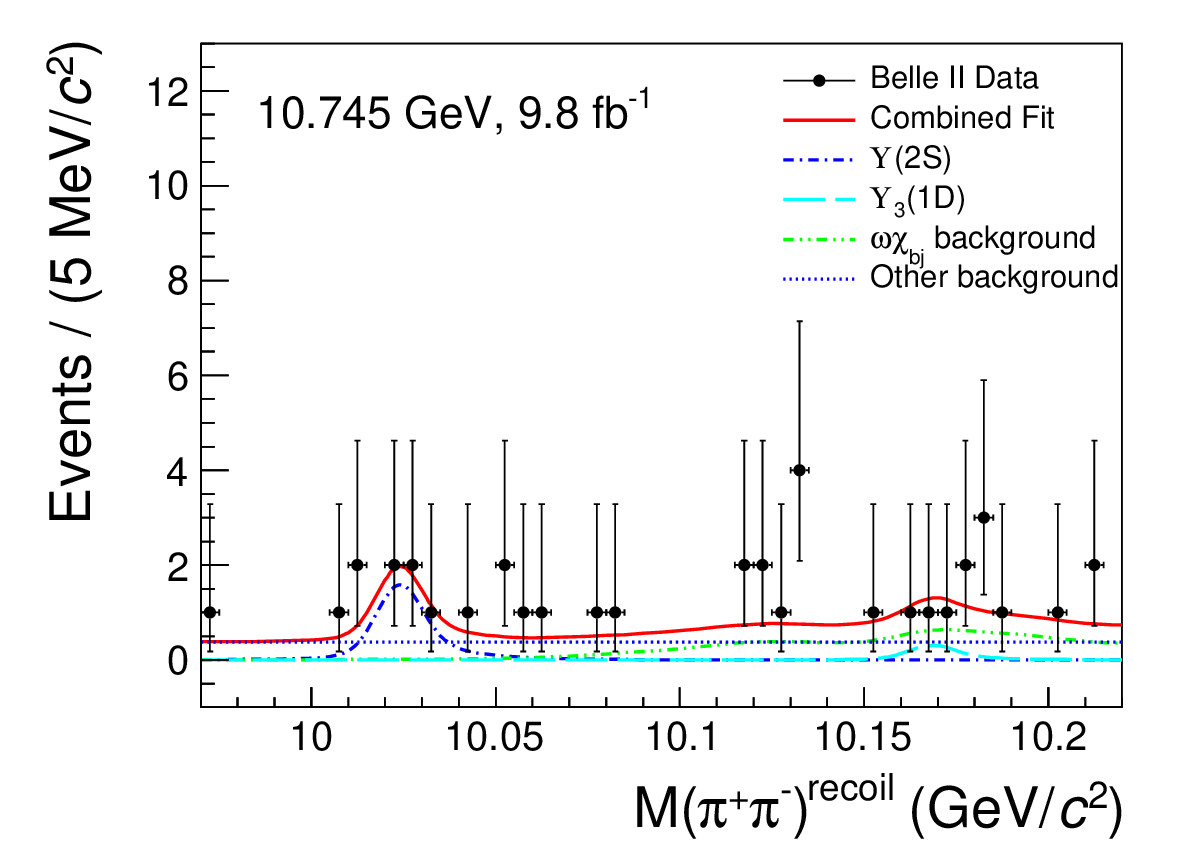}
  \includegraphics[height=2.1in]{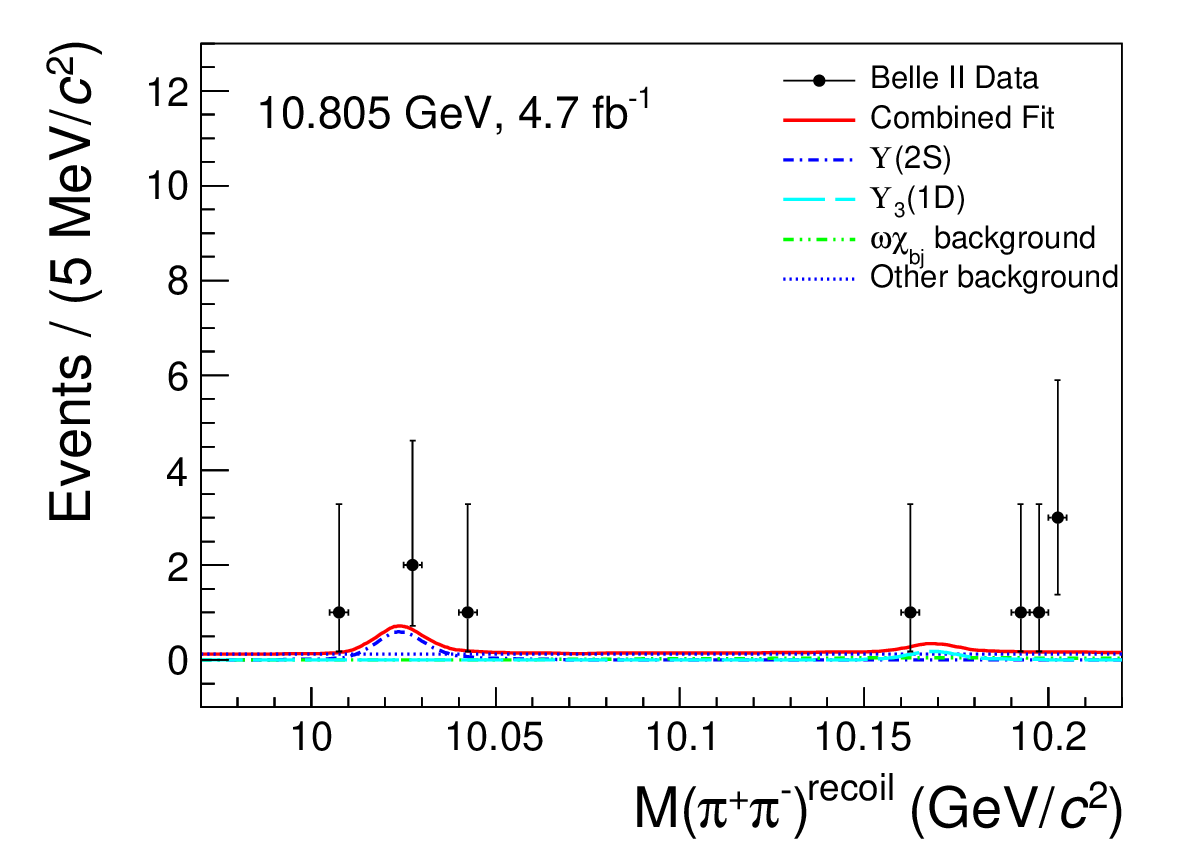}
  \caption{Fits to the \( M(\pi^+\pi^-)^{\text{recoil}} \) distributions for data events in the \( \chi_{b2} \) channel. 
  Dots with error bars represent the data, the solid red line is the total fit, the long-dashed cyan line is the \( \Upsilon_3(1D) \) signal component, the dot-dashed blue line is the \( \Upsilon(2S) \) signal component, the double-dot-dashed green line represents the \( \omega \chi_{b1,b2} \) background, and the dotted blue line is the constant background component.
  Note that the \(\sqrt{s} = 10.701~\mathrm{GeV}\) sample contains zero events.
  }
  \label{pipiRecoilData-chib2}
\end{figure}

The \(M(\pi^+\pi^-)^{\rm recoil}\) distributions for signal MC samples, illustrating the expected shapes for \(\Upsilon_2(1D)\), \(\Upsilon_3(1D)\), and \(\Upsilon(2S)\), are shown in Fig.~\ref{pipiRecoilMC}, normalized to unit area for clarity.
Figures~\ref{pipiRecoilData} and \ref{pipiRecoilData-chib2} present the \(M(\pi^+\pi^-)^{\rm recoil}\) distributions for data events at each center-of-mass energy after applying all selection criteria, corresponding to the \(\chi_{b1}\) and \(\chi_{b2}\) channels, respectively. 
No significant \(\Upsilon_J(1D)\) signal is observed at any of the collision energies studied.
A small number of \(\Upsilon(2S)\) events are observed at \(\sqrt{s} = 10.745\) and \(10.805\)~GeV, as expected, originating from the process \(e^+e^- \to \pi^+\pi^- \Upsilon(2S)\). The measured cross sections for this control process agree with prior Belle and Belle~II results within uncertainties~\cite{Belle:pipiYnS2008,Belle:pipiYnS,Belle-II:pipiYnS}; 
we do not report the obtained numerical values here, as the statistical precision is insufficient to impact the averages of those previous measurements.

Signal yields are extracted via an unbinned extended maximum likelihood fit to each distribution. The fitted components are shown in Figs.~\ref{pipiRecoilData} and \ref{pipiRecoilData-chib2}.
For the data sample at \(\sqrt{s} = 10.701\)~GeV which contains zero events, a binned fit with a uniform bin width of 5~MeV is employed instead.
In this case, the number of entries in each bin is treated as an independent Poisson observable, and the likelihood function is constructed accordingly.
The probability density functions (PDFs) for the \(\Upsilon(2S)\) and \(\Upsilon_J(1D)\) signals are parameterized using shapes derived from their respective MC simulations (MC histograms). The yields for the \(\Upsilon(2S)\) and \(\Upsilon_J(1D)\) signals are allowed to float freely in the fit.
To mitigate potential anticorrelations between the fitted yields of the \(\Upsilon_2(1D)\) and \(\Upsilon_3(1D)\) states, which arise from their small mass difference and the resulting overlap of their signal shapes, we neglect cross-feed by fixing the contribution from the alternate state to zero.
Specifically, for the $  \chi_{b1}  $ channel, we set the $  \Upsilon_3(1D)  $ yield to zero when extracting $  \Upsilon_2(1D)  $, and vice versa for the $  \chi_{b2}  $ channel. This approach ensures fit stability and yields more conservative upper limits.
As a cross-check, we tested the impact of possible cross-feed between the \(\Upsilon_2(1D)\) and \(\Upsilon_3(1D)\) states by allowing both yields to float simultaneously in the fit; the resulting upper limits changed negligibly.
The background PDF consists of a constant term, with its yield allowed to float freely in the fit, and the \(e^+e^- \to \omega \chi_{b1,b2}\) component. The \(\omega \chi_{b1,b2}\) contribution, shaped from MC simulations and normalized to the cross sections reported in Ref.~\cite{belle2-omegachibj}, is fixed during the fit. Notably, this background vanishes at \(\sqrt{s} = 10.653\) and \(10.701\)~GeV due to negligibly small production cross sections at these energies~\cite{belle2-omegachibj}.

To validate the reliability of the fitting procedure and the upper limit construction in this low-statistics regime, we performed an ensemble test using MC pseudo-experiments. We generate 2000 signal-plus-background pseudo-experiments and 2000 background-only pseudo-experiments according to the fit model employed above. Each pseudo-experiment is fitted with a model configuration identical to what is used for real data. The pull distributions of the fitted signal yields are well described by a Gaussian function, yielding a mean of \(0.003 \pm 0.018\) (\(-0.013 \pm 0.018\)) and a width of \(0.997 \pm 0.013\) (\(0.994 \pm 0.013\)) under the signal-plus-background (background-only) hypothesis. The zero mean and unit width confirm that the fit is unbiased and performs as expected.

\section{Cross Section Measurement}

Given the absence of a significant signal for \(\Upsilon_J(1D)\) in the experimental data, we employ a Bayesian approach~\cite{pdg} to calculate the upper limit on the signal yield at 90\% credibility for each center-of-mass energy. This method involves integrating the likelihood function over the signal yield parameter, ranging from zero to infinity, and determining the upper limit as the value where the cumulative probability reaches 90\% of the total. By excluding negative signal yields, this approach ensures a conservative estimation.
To account for systematic uncertainties, we convolve the likelihood function with a Gaussian function centered at each assumed signal yield, with the width of the Gaussian corresponding to the total systematic uncertainty. 

At each center-of-mass energy, the upper limit on the signal yield of \(\Upsilon_J(1D)\) is subsequently converted into an upper limit on the product of the dressed cross section and branching fraction for the specific decay channels $e^+e^- \to \pi^+\pi^- \Upsilon_2(1D)$ with $\Upsilon_2(1D) \to \gamma \chi_{b1}$, and $e^+e^- \to \pi^+\pi^- \Upsilon_3(1D)$ with $\Upsilon_3(1D) \to \gamma \chi_{b2}$. The calculation is performed using the following expression:
\begin{equation}
\begin{aligned}
&\sigma [e^+e^- \to \pi^+\pi^- \Upsilon_2(1D)] \times \mathcal{B}[\Upsilon_2(1D) \to \gamma \chi_{b1}] \\
&= \frac{N_{\Upsilon_2(1D)} (1 - R_{\chi_{b2}})}{\mathcal{L} \mathcal{B}[\chi_{b1} \to \gamma \Upsilon(1S)] \mathcal{B}[\Upsilon(1S) \to \ell^+\ell^-] \epsilon^{\rm w} (1 + \delta)},
\end{aligned}
\end{equation}
and
\begin{equation}
\begin{aligned}
&\sigma [e^+e^- \to \pi^+\pi^- \Upsilon_3(1D)] \times \mathcal{B}[\Upsilon_3(1D) \to \gamma \chi_{b2}] \\
&= \frac{N_{\Upsilon_3(1D)}}{\mathcal{L} \mathcal{B}[\chi_{b2} \to \gamma \Upsilon(1S)] \mathcal{B}[\Upsilon(1S) \to \ell^+\ell^-] \epsilon^{\rm w} (1 + \delta)}.
\end{aligned}
\end{equation}
Here, $N_{\Upsilon_J(1D)}$ represents the upper limit on the number of signal events for the respective $\Upsilon_J(1D)$ channel, $\mathcal{L}$ is the integrated luminosity, and $\mathcal{B}$ denotes the relevant branching fractions.

MC simulations reveal an overlap between the $\chi_{b1}$ and $\chi_{b2}$ peaks, resulting in potential cross-contamination between them. 
For the $\Upsilon_2(1D)$ channel, the contamination from the $\Upsilon_2(1D) \to \gamma \chi_{b2}$ decay is quantified by the ratio $R_{\chi_{b2}}$, which is computed as:
\begin{equation}
R_{\chi_{b2}} =
\frac{
  \epsilon^{\Upsilon_2(1D)\to\gamma\chi_{b2}}\,
  \mathcal{B}[\Upsilon_2(1D)\to\gamma\chi_{b2}]\,
  \mathcal{B}[\chi_{b2}\to\gamma\Upsilon(1S)]
}{
  \sum_{i=1}^{2}\,
  \epsilon^{\Upsilon_2(1D)\to\gamma\chi_{bi}}\,
  \mathcal{B}[\Upsilon_2(1D)\to\gamma\chi_{bi}]\,
  \mathcal{B}[\chi_{bi}\to\gamma\Upsilon(1S)]
},
\label{eq:Rchi}
\end{equation}
where the selection efficiencies \(\epsilon^{\Upsilon_2(1D)\to\gamma\chi_{bi}}\) (\(i=1,2\)) are determined directly from the corresponding signal MC samples, the branching fractions \(\mathcal{B}[\Upsilon_2(1D)\to\gamma\chi_{bi}]\) are taken from the theoretical predictions of Ref.~\cite{yjd-BR}, and the remaining branching fractions \(\mathcal{B}[\chi_{bi}\to\gamma\Upsilon(1S)]\) are taken from the PDG~\cite{pdg}. Conversely, for the $\Upsilon_3(1D)$ channel, the contamination from $\Upsilon_3(1D) \to \gamma \chi_{b1}$ is negligible, as the branching fraction for this decay is predicted to be nearly zero according to theoretical calculations~\cite{yjd-BR}.

The re-weighted reconstruction efficiency, $\epsilon^{\rm w}$, is obtained by adjusting the $\frac{1}{s}$ cross section line-shape used in the MC generation to match the $\Upsilon(10753)$ resonance line-shape, following the methodology outlined in Refs.~\cite{Belle-II:pipiYnS,ISRweighting}. The radiative correction factor, $(1 + \delta)$, is defined as:

\begin{equation}
(1 + \delta) = \frac{\int_0^{x_m} \sigma^{\rm dress}(s(1 - x)) W(x, s) \, dx}{\sigma^{\rm dress}(s)},
\end{equation}
where $\sigma^{\rm dress}(s)$ is the energy-dependent dressed cross section modeled using the $\Upsilon(10753)$ resonance line-shape. The radiator function $W(x, s)$ accounts for ISR effects, as described in Ref.~\cite{radiator}. The upper integration limit, $x_m = 1 - s_m/s$, represents the maximum fraction of energy carried by the ISR photon, with $s_m = [m(\pi^+) + m(\pi^-) + m[\Upsilon_J(1D)]]^2$ being the minimum energy squared required to produce the final state.

Tables~\ref{cs-measurement-y1d} and~\ref{cs-measurement-y31d} present the measurements of \(\sigma[e^+e^- \to \pi^+\pi^- \Upsilon_2(1D)] \times \mathcal{B}[\Upsilon_2(1D) \to \gamma \chi_{b1}]\) and \(\sigma[e^+e^- \to \pi^+\pi^- \Upsilon_3(1D)] \times \mathcal{B}[\Upsilon_3(1D) \to \gamma \chi_{b2}]\), respectively, assuming the \(\Upsilon(10753)\) resonance line-shape hypothesis for the cross section.
To offer an alternative perspective, the same tables include values in parentheses, corresponding to estimates derived under the \(\Upsilon(5S)\) resonance hypothesis, where the radiative correction factor $(1+\delta)$ and the efficiency $\epsilon^{\rm w}$ are recalculated to reflect the $\Upsilon(5S)$ line-shape.
It is noteworthy that the center-of-mass energies of the data samples in this analysis (\(\sqrt{s} = 10.653\), 10.701, 10.745, and 10.805 GeV) lie well below the \(\Upsilon(5S)\) peak, where its contribution is expected to be minimal, in contrast to the \(\Upsilon(10753)\) resonance, where these energies cover the peak region of the resonance.

\begin{table}
    \begin{center}
    \caption{
    Measurements on the product of the dressed cross section and branching fraction, \(\sigma[e^+e^- \to \pi^+\pi^- \Upsilon_2(1D)] \times \mathcal{B}[\Upsilon_2(1D) \to \gamma \chi_{b1}]\), at various center-of-mass energies \(\sqrt{s}\) (in GeV). 
    Parameters include integrated luminosity \(\mathcal{L}\) (in fb$^{-1}$), re-weighted efficiency \(\epsilon^{\text{w}}\), radiative correction factor \((1 + \delta)\), signal yield \(N_{\Upsilon_2(1D)}\), contamination ratio \(R_{\chi_{b2}}\), and the resulting product \(\sigma \times \mathcal{B}\) (in pb, incorporating systematic uncertainties), estimated under the \(\Upsilon(10753)\) line-shape hypothesis for the cross section.
    Values in the parentheses are for the \(\Upsilon(5S)\) line-shape hypothesis.
    }
    \label{cs-measurement-y1d}
    \begin{tabular}{ccccccc}
      \hline\hline
      $\sqrt{s}$ & $\mathcal{L}$ & $\epsilon^{\text{w}}$ & $(1+\delta)$ & $N_{\Upsilon_2(1D)}$ & $R_{\chi_{b2}}$ & $\sigma \times \mathcal{B}$ (pb) \\
      \hline
      10.653 & 3.5 & 0.180 (0.172) & 0.895 (0.819) & $<4.6$ & 0.099 & $<0.43$ ($<0.49$) \\
      10.701 & 1.6 & 0.202 (0.177) & 0.723 (0.923) & $<2.3$ & 0.100 & $<0.52$ ($<0.46$) \\
      10.745 & 9.8 & 0.215 (0.191) & 0.587 (0.885) & $<6.8$ & 0.099 & $<0.29$ ($<0.21$) \\
      10.805 & 4.7 & 0.176 (0.216) & 1.039 (0.732) & $<3.0$ & 0.100 & $<0.18$ ($<0.21$) \\
      \hline\hline
    \end{tabular}
    \end{center}
\end{table}

\begin{table}
    \begin{center}
    \caption{
    Measurements on the product of the dressed cross section and branching fraction, \(\sigma[e^+e^- \to \pi^+\pi^- \Upsilon_3(1D)] \times \mathcal{B}[\Upsilon_3(1D) \to \gamma \chi_{b2}]\), at various center-of-mass energies \(\sqrt{s}\) (in GeV).
    Parameters include integrated luminosity \(\mathcal{L}\) (in fb$^{-1}$), re-weighted efficiency \(\epsilon^{\text{w}}\), radiative correction factor \((1 + \delta)\), signal yield \(N_{\Upsilon_3(1D)}\), and the resulting product \(\sigma \times \mathcal{B}\) (in pb, incorporating systematic uncertainties), estimated under the \(\Upsilon(10753)\) line-shape hypothesis for the cross section.
    Values in the parentheses are for the \(\Upsilon(5S)\) line-shape hypothesis.
    }
    \label{cs-measurement-y31d}
    \begin{tabular}{cccccc}
      \hline\hline
      $\sqrt{s}$ & $\mathcal{L}$ & $\epsilon^{\text{w}}$ & $(1+\delta)$ & $N_{\Upsilon_3(1D)}$ & $\sigma \times \mathcal{B}$ (pb) \\
      \hline
      10.653 & 3.5 & 0.177 (0.169) & 0.893 (0.827) & $<3.6$ & $<0.73$ ($<0.83$) \\
      10.701 & 1.6 & 0.202 (0.180) & 0.724 (0.901) & $<2.3$ & $<1.12$ ($<1.01$) \\
      10.745 & 9.8 & 0.212 (0.194) & 0.587 (0.883) & $<7.4$ & $<0.69$ ($<0.53$) \\
      10.805 & 4.7 & 0.170 (0.209) & 1.042 (0.728) & $<5.2$ & $<0.71$ ($<0.83$) \\
      \hline\hline
    \end{tabular}
    \end{center}
\end{table}

\section{Systematic Uncertainty}

The systematic uncertainties in the cross section measurement stem from several sources. These include uncertainties in integrated luminosity, trigger simulation, tracking efficiency, photon detection efficiency, branching fractions, decay model, MC statistics, signal parameterization, and background modeling.
Each contributing factor is elaborated below.

The integrated luminosity at Belle II is determined with a precision of \(0.6\%\) using Bhabha, di-photon and di-muon events, as documented in Ref.~\cite{lum-belleii}.
The systematic uncertainty associated with the trigger simulation is conservatively evaluated at \(1.0\%\). This estimate is based on discrepancies in trigger efficiencies between data (99.9\%) and MC simulations (99.3\%), analyzed through the \(e^+e^- \to 2(\pi^+\pi^-\pi^0)\) control sample.
The reconstruction efficiency of a single high-momentum charged track ($p > 0.2$~GeV/$c$) carries a systematic uncertainty of \(0.3\%\), derived from tracking efficiency comparisons between data and MC using \(\tau\)-pair events; for the two high-momentum leptons in the signal, this yields a total uncertainty of $0.6\%$.
For pion tracks, the reconstruction efficiency uncertainty is 0.3\% per track at high momentum ($p > 0.2$~GeV/$c$), while efficiencies for low-momentum pions ($p < 0.2$~GeV/$c$) in the signal MC are reweighted to match data based on studies of $B^0 \to D^{*-}\pi^+$ decays; the combined uncertainty for the pion tracks is estimated at 0.9\%.
Photon detection efficiencies in the signal MC are adjusted to align with data, using the data-to-simulation ratios of photon efficiency measured with \(e^+e^- \to \mu^+\mu^-\gamma_{\rm ISR}\) events.
These ratios are provided in bins of the photon polar and azimuthal angles over the energy range \([0.2, 7]\) GeV, which substantially covers the photon energies of the signal process (approximately 100--600 MeV).
Photons below 200 MeV constitute only about 1\% of the total and are not corrected; their contribution to the overall uncertainty is negligible.
The associated systematic uncertainty on the photon efficiency, derived from the uncertainties on these correction ratios, is \(3.6\%\).

The branching fractions \(\mathcal{B}[\chi_{b1} \to \gamma \Upsilon(1S)]\), \(\mathcal{B}[\chi_{b2} \to \gamma \Upsilon(1S)]\), \(\mathcal{B}[\Upsilon(1S) \to e^+e^-]\), and \(\mathcal{B}[\Upsilon(1S) \to \mu^+\mu^-]\) have uncertainties of \(5.7\%\), \(5.6\%\), \(3.3\%\), and \(1.6\%\), respectively, as given by the PDG~\cite{pdg}. These uncertainties are treated as uncorrelated for the \(\chi_{b1}\) and \(\chi_{b2}\) channels. The combined systematic uncertainties for the product \(\mathcal{B}[\chi_{bJ'} \to \gamma \Upsilon(1S)] \mathcal{B}[\Upsilon(1S) \to \ell^+\ell^-]\) are computed as \(6.3\%\) for both \(J' = 1\) and \(J' = 2\).

The factor \((1 - R_{\chi_{b2}})\) is applied to correct for the small contamination from \(\Upsilon_2(1D) \to \gamma \chi_{b2}\) decays that are reconstructed in the \(\chi_{b1}\) selection. Owing to the nearly identical final state topologies and kinematics of the two decay chains, the reconstruction efficiencies \(\epsilon^{\Upsilon_2(1D)\to\gamma\chi_{b1}}\) and \(\epsilon^{\Upsilon_2(1D)\to\gamma\chi_{b2}}\) are highly correlated; their uncertainties therefore cancel in the ratio \(R_{\chi_{b2}}\). The theoretical branching fractions \(\mathcal{B}[\Upsilon_2(1D)\to\gamma\chi_{b1/b2}]\), taken from the same model~\cite{yjd-BR}, are likewise strongly correlated, and their uncertainties also cancel.
Furthermore, independent theoretical calculations~\cite{Segovia:2016xqb,Wang:2018rjg,Godfrey:2015dia} predict similar values for these branching fractions, yielding identical results for the ratio.
Consequently, propagating only the uncertainties on \(\mathcal{B}[\chi_{b1} \to \gamma \Upsilon(1S)]\) (5.7\%) and \(\mathcal{B}[\chi_{b2} \to \gamma \Upsilon(1S)]\) (5.6\%)~\cite{pdg} yields a relative uncertainty of \(0.8\%\) on \((1 - R_{\chi_{b2}})\).

The default signal MC simulation employs a three-body PHSP model for \(\pi^+\pi^-\Upsilon_J(1D)\) production. To evaluate the systematic uncertainty associated with the \(\pi^+\pi^-\) dynamics, an alternative model incorporating the scalar resonance \(f_0(500)\) (\(J^{PC} = 0^{++}\)) is used. 
This resonance is the most relevant contribution in the kinematically allowed \(\pi^+\pi^-\) invariant mass region of lower than 0.6~GeV in this analysis.
(The possible resonance \(f_0(980)\) is also examined and contributes only marginally due to its higher mass.
)
The relative difference in reconstruction efficiency between the default PHSP and the \(f_0(500)\) models is 3.6\%, which is assigned as the systematic uncertainty for the decay model.
The estimate also accounts for effects from the \(\gamma\)-conversion background veto applied to \(\cos(\theta_{\pi^+\pi^-})\).

The systematic uncertainty due to the size of the MC sample is determined using the expression \(\sqrt{\frac{\epsilon (1 - \epsilon)}{N}} / \epsilon\), where \(\epsilon\) represents the signal reconstruction efficiency and \(N\) is the number of generated MC events. This uncertainty is estimated to be \(0.4\%\).

Direct comparisons of fits to data can be significantly affected by statistical fluctuations, potentially masking true systematic effects. To address this, we conduct a MC pseudo-experiment study to quantify the systematic uncertainty associated with signal parameterization. We generate 2000 MC pseudo-experiment samples using the default signal PDF, derived from the signal MC shape. Each sample is fitted with two signal models: the default signal PDF and an alternative model defined as a Crystal Ball function~\cite{Oreglia:1980cs}.
To account for possible differences in resolution between data and MC simulation, the resolution parameter of the Crystal Ball function is first adjusted to match the signal MC shape and then varied within \(\pm 2.2\)~MeV.
This \(\pm 2.2\)~MeV range corresponds to the resolution difference between data and MC measured from the control sample \(e^+e^- \to \pi^+\pi^-\Upsilon(2S)[\to \ell^+\ell^-]\)~\cite{Belle-II:pipiYnS}, where the difference is determined to be (\(0.1 \pm 2.2\))~MeV.
The resulting relative difference in fitted signal yield between the two models is 0.5\%, which is assigned as the systematic uncertainty due to signal parameterization.

We assess the systematic uncertainty associated with the background modeling using a similar MC pseudo-experiment approach. We generate 2000 MC pseudo-experiment samples, each containing signal and background events. The signal is modeled with the default signal PDF, while the background is modeled with the default 0th-order polynomial (constant term). Each sample is fitted with two background models: the default model and an alternative model defined as a first-order polynomial. By comparing the signal yield distributions from the two fits, we estimate the impact of varying the background PDF shape. The relative difference in signal yield is calculated to be 0.2\%, which is assigned as the systematic uncertainty due to the background shape.

The radiative correction in \(e^+e^-\) collisions depends on the line-shape of the cross section for the process \(e^+e^- \to \pi^+\pi^- \Upsilon_J(1D)\). Due to uncertainty in the exact line-shape, we refrain from assigning a systematic uncertainty to the radiative correction. Instead, we adopt an approach where the cross section is evaluated under two distinct hypotheses, as mentioned earlier: one assuming it follows the \(\Upsilon(10753)\) resonance line-shape, and another assuming the \(\Upsilon(5S)\) resonance line-shape. Results based on both hypotheses are reported in Tables~\ref{cs-measurement-y1d} and~\ref{cs-measurement-y31d}. 

The systematic uncertainty from the \(\Upsilon(1S)\) signal window on the \(M(\ell^+\ell^-)\) invariant mass is negligible, due to the accurate modeling of lepton resolution in the MC. This is supported by lepton performance studies using \(\tau\)-pair events, as well as resolution investigations in the control sample \(e^+e^- \to \pi^+\pi^- \Upsilon(2S) \to \pi^+\pi^- \ell^+\ell^-\)~\cite{Belle-II:pipiYnS}. The systematic uncertainty from the \(\chi_{b1,b2}\) signal window on the \(M[\gamma_H \Upsilon(1S)]\) invariant mass is negligible due to the photon energy corrections applied to the MC simulation to match experimental data, using the \(\pi^0 \to \gamma\gamma\) sample. The systematic uncertainty from the \(\eta\) background veto on \(M(\gamma_L \gamma_H)\) invariant mass is negligible, as it is mitigated by the same photon energy corrections.

Assuming all systematic uncertainty sources are independent, the total systematic uncertainty is obtained by combining them in quadrature, resulting in a value of $8.3\%$. A summary of the systematic uncertainties is presented in Table~\ref{sec-err}.

\begin{table}[h]
\begin{center}
\caption{Systematic uncertainties for the measurement of \(\sigma [e^+e^- \to \pi^+\pi^- \Upsilon_J(1D)] \times \mathcal{B}[\Upsilon_J(1D) \to \gamma \chi_{bJ'}]\).}
\label{sec-err}
\begin{tabular}{lc}
  \hline\hline
  Source                    & Uncertainty (\%)        \\
  \hline
  Integrated luminosity     & 0.6              \\
  Trigger simulation        & 1.0              \\
  Lepton efficiency         & 0.6              \\
  Pion efficiency           & 0.9              \\
  Photon efficiency         & 3.6              \\
  Branching fractions       & 6.3              \\
  \(1-R_{\chi_{b2}}\)       & 0.8              \\
  Decay model               & 3.6              \\
  MC sample size            & 0.4              \\
  Signal parameterization   & 0.5              \\
  Background shape          & 0.2              \\
  \hline
  Quadrature sum            & $8.3$              \\
  \hline\hline
\end{tabular}
\end{center}
\end{table}

The systematic uncertainty associated with the fixed background from \(e^+e^- \to \omega \chi_{b1,b2}\) in the signal extraction is evaluated by varying its normalization by \(\pm 24\%\), corresponding to the combined statistical and systematic uncertainty reported in Ref.~\cite{belle2-omegachibj}. For each variation, the upper limit on the \(\Upsilon_J(1D)\) signal yield is recalculated, and the most conservative value, i.e.\ the largest upper limit obtained from the nominal and varied background scenarios, is adopted as the final result.
This adjustment increases the upper limits on the \(\Upsilon_J(1D)\) signal yields by 0 to 0.7 events, with the following energy-dependent impacts: the \(\Upsilon_2(1D)\) (\(\Upsilon_3(1D)\)) yield upper limit rises from 6.2 (6.7) to 6.8 (7.4) at \(\sqrt{s} = 10.745\)~GeV and from 3.0 (5.1) to 3.0 (5.2) at \(\sqrt{s} = 10.805\)~GeV, while no change occurs at \(\sqrt{s} = 10.653\) and \(10.701\)~GeV where this background component is negligible.

\section{Discussion}

\begin{figure}[h]
  \centering
  \includegraphics[height=2.1in]{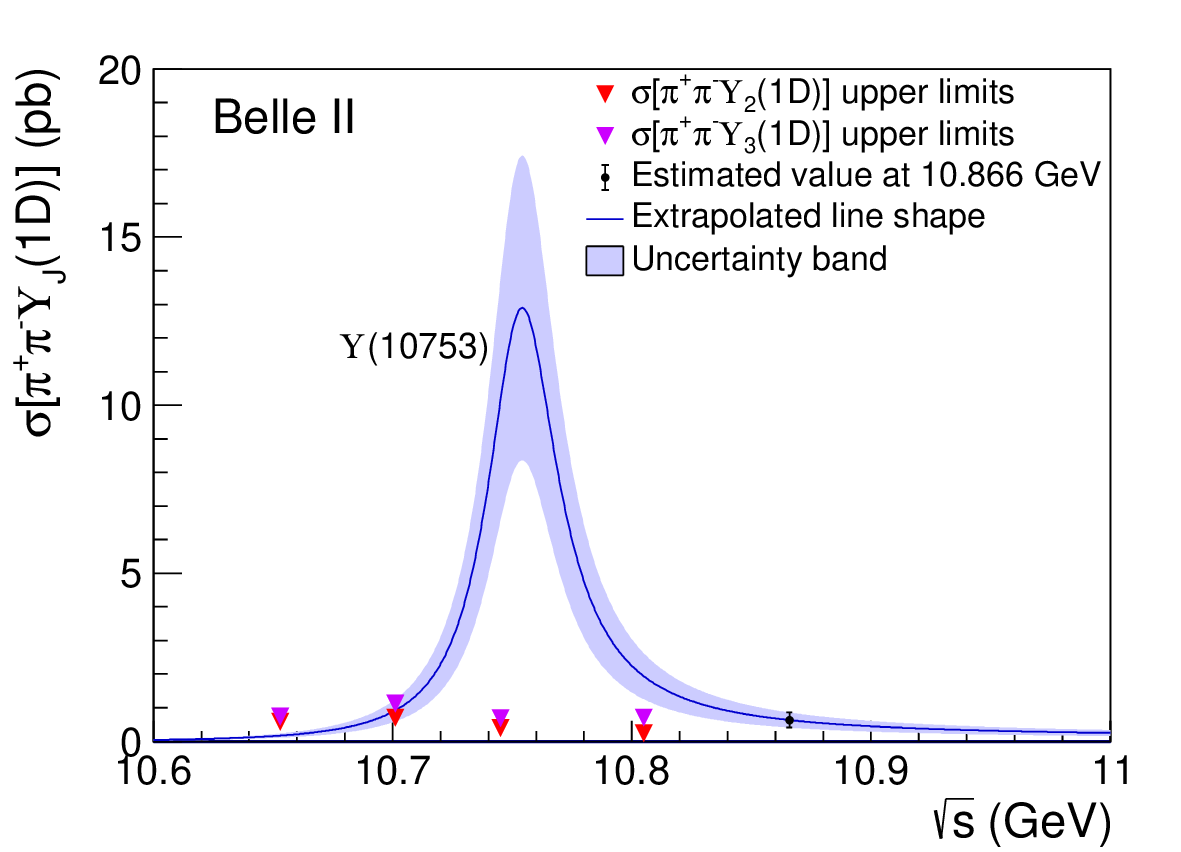}
  \includegraphics[height=2.1in]{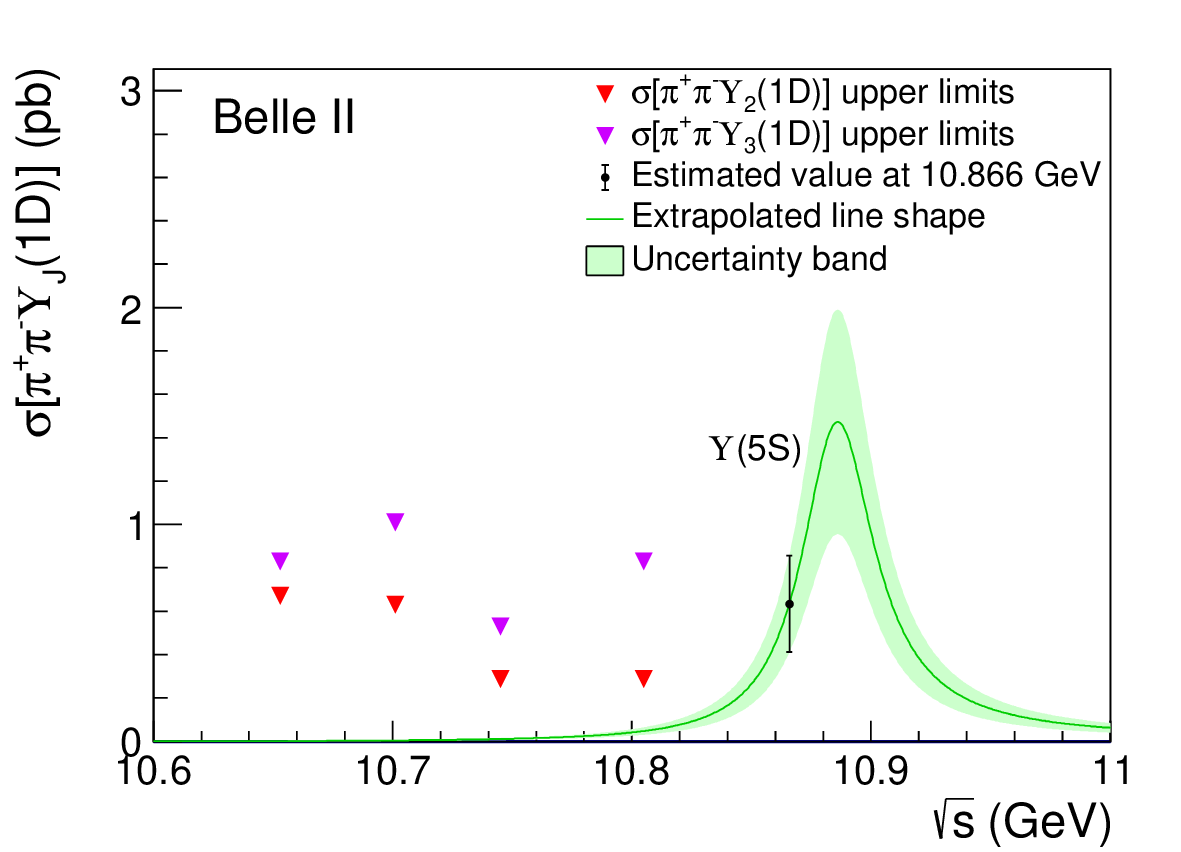}
  \caption{
  The 90\% credibility upper limits on the \(\sigma[e^+e^- \to \pi^+\pi^- \Upsilon_J(1D)]\) as a function of center-of-mass energy, evaluated under the \(\Upsilon(10753)\) resonance hypothesis (left) and the \(\Upsilon(5S)\) resonance hypothesis (right). Inverted triangles denote the estimated upper limits.
  The black dots at \(\sqrt{s} = 10.866\) GeV indicate the estimated value from Belle results~\cite{y5s-pipiInclusive-belle}, with error bars representing the combined statistical and systematic uncertainties.
  The solid curves represent the extrapolated cross section line shapes under the respective resonance hypotheses, with the filled bands illustrating the uncertainties propagated from the Belle measurement while following the line-shape of each resonance.
  }
  \label{fig:cross-section}
\end{figure}

In an inclusive analysis of \(\Upsilon(5S) \to \pi^+ \pi^- X\) decays at Belle~\cite{y5s-pipiInclusive-belle}, the yields derived from fits to the dipion missing mass spectrum are \((143.8 \pm 8.7 \pm 6.8) \times 10^3\) and \((22.4 \pm 7.8) \times 10^3\) for \(\Upsilon(2S)\) and \(\Upsilon_J(1D)\), respectively.
(The uncertainty on the \(\Upsilon_J(1D)\) yield is statistical only; the Belle Collaboration notes that the \(\Upsilon_J(1D)\) signal is marginal and the associated systematic uncertainty is therefore negligible compared to the statistical component~\cite{y5s-pipiInclusive-belle}.)
Utilizing the cross section for \(e^+e^- \to \pi^+\pi^- \Upsilon(2S)\) at 10.866 GeV, measured as \(4.07 \pm 0.18\) pb in Ref.~\cite{Belle:pipiYnS} with the same data sample, we estimate the cross section for \(e^+e^- \to \pi^+\pi^- \Upsilon_J(1D)\) to be \(0.63 \pm 0.22\) pb at this energy. This estimation assumes comparable reconstruction efficiencies between \(\Upsilon(2S)\) and \(\Upsilon_J(1D)\).
Starting from this estimated value, we extrapolate the energy-dependent cross sections \(\sigma[e^+e^- \to \pi^+\pi^- \Upsilon_J(1D)]\) under two distinct hypotheses: production from the \(\Upsilon(10753)\) resonance and production from the \(\Upsilon(5S)\) resonance (a Breit-Wigner line-shape with parameters taken from the PDG~\cite{pdg} is assumed for each resonance).
These extrapolated cross sections are illustrated in Fig.~\ref{fig:cross-section} (solid curves with uncertainty bands), with the left panel depicting the \(\Upsilon(10753)\) hypothesis and the right panel the \(\Upsilon(5S)\) hypothesis.
We note that the current dataset has limited statistics, and covers only the energy region around the \(\Upsilon(10753)\) without spanning the full range of both the \(\Upsilon(10753)\) and \(\Upsilon(5S)\) resonances. Consequently, only the above two limiting hypotheses can be tested; the scenario of combined contributions from both resonances cannot yet be constrained.

Our analysis provides 90\% credibility upper limits on the products \(\sigma[e^+e^- \to \pi^+\pi^-\Upsilon_{2(3)}(1D)] \times \mathcal{B}[\Upsilon_{2(3)}(1D) \to \gamma \chi_{b1(2)}]\) (Tables~\ref{cs-measurement-y1d} and~\ref{cs-measurement-y31d}).
To enable a direct comparison with the extrapolated cross sections, these limits are divided by the theoretical branching fractions \(\mathcal{B}(\Upsilon_2(1D) \to \gamma \chi_{b1}) \approx 0.73\) and \(\mathcal{B}(\Upsilon_3(1D) \to \gamma \chi_{b2}) \approx 1\)~\cite{yjd-BR,Segovia:2016xqb,Wang:2018rjg,Godfrey:2015dia}, respectively. The uncertainties on these theoretical branching fractions (4\%) are negligible compared to the dominant statistical uncertainty on the extrapolated cross sections (35\%).
The resulting upper limits are displayed as inverted triangles in Fig.~\ref{fig:cross-section}.

We assume that the \(\pi^+\pi^-\Upsilon_J(1D)\) system observed by Belle~\cite{y5s-pipiInclusive-belle} is dominated by the \(J=2\) component. This assumption is motivated by the conclusive identification of the \(\Upsilon_2(1D)\) (\(1^3D_2\)) state by the CLEO~\cite{CLEO:2004npj} and BaBar~\cite{BaBar:2010tqb} collaborations (which reported inconclusive evidence for other \(J\) states) and the similar reported mass of \(\Upsilon_J(1D)\).
Our measurement reveals that the \(\pi^+\pi^- \Upsilon_J(1D)\) system is more consistent with production from the \(\Upsilon(5S)\) resonance, as the estimated upper limits are well above the expected values, essentially null, under this hypothesis (Fig.~\ref{fig:cross-section}, right panel). Conversely, at \(\sqrt{s} = 10.745\) GeV, near the \(\Upsilon(10753)\) peak, the upper limits fall significantly below the anticipated cross section (Fig.~\ref{fig:cross-section}, left panel).
Consequently, the estimated upper limits do not favor a scenario in which the \(\pi^+\pi^-\Upsilon_J(1D)\) system is produced solely from the \(\Upsilon(10753)\) resonance.
The same conclusion can be reached in the case that the system is dominated by the \(\Upsilon_3(1D)\) (\(J=3\)) component.
These findings, when considered alongside the \(\Upsilon(5S)\) resonance, contribute to the elucidation of the \(\Upsilon(10753)\)'s intrinsic nature.

\section{Summary}

We present a comprehensive study of the processes \( e^+e^- \to \pi^+\pi^- \Upsilon_J(1D) \), where \( J = 2 \) or \( 3 \), utilizing a 19.6~fb\(^{-1}\) data sample collected by the Belle II detector at the SuperKEKB \( e^+e^- \) asymmetric-energy collider. The data were acquired at center-of-mass energies of \( \sqrt{s} = 10.653 \), \( 10.701 \), \( 10.745 \), and \( 10.805 \)~GeV, strategically chosen to probe the \(\Upsilon(10753)\) resonance.
For each center-of-mass energy, upper limits at 90\% credibility on the products of the cross sections and branching fractions, \(\sigma[e^+e^- \to \pi^+\pi^- \Upsilon_2(1D)] \times \mathcal{B}(\Upsilon_2(1D) \to \gamma \chi_{b1})\) and \(\sigma[e^+e^- \to \pi^+\pi^- \Upsilon_3(1D)] \times \mathcal{B}(\Upsilon_3(1D) \to \gamma \chi_{b2})\), are determined and reported.

\acknowledgments

This work, based on data collected using the Belle II detector, which was built and commissioned prior to March 2019,
was supported by
Higher Education and Science Committee of the Republic of Armenia Grant No.~23LCG-1C011;
Australian Research Council and Research Grants
No.~DP200101792, 
No.~DP210101900, 
No.~DP210102831, 
No.~DE220100462, 
No.~LE210100098, 
and
No.~LE230100085; 
Austrian Federal Ministry of Education, Science and Research,
Austrian Science Fund (FWF) Grants
DOI:~10.55776/P34529,
DOI:~10.55776/J4731,
DOI:~10.55776/J4625,
DOI:~10.55776/M3153,
and
DOI:~10.55776/PAT1836324,
and
Horizon 2020 ERC Starting Grant No.~947006 ``InterLeptons'';
Natural Sciences and Engineering Research Council of Canada, Digital Research Alliance of Canada, and Canada Foundation for Innovation;
National Key R\&D Program of China under Contract No.~2024YFA1610503,
and
No.~2024YFA1610504
National Natural Science Foundation of China and Research Grants
No.~11575017,
No.~11761141009,
No.~11705209,
No.~11975076,
No.~12135005,
No.~12150004,
No.~12161141008,
No.~12405099,
No.~12475093,
and
No.~12175041,
Shandong Provincial Natural Science Foundation Project~ZR2022JQ02,
and Shandong Postdoctoral Science Foundation under Contract No.~SDCX-ZG-202400324;
the Czech Science Foundation Grant No. 22-18469S,  Regional funds of EU/MEYS: OPJAK
FORTE CZ.02.01.01/00/22\_008/0004632 
and
Charles University Grant Agency project No. 246122;
European Research Council, Seventh Framework PIEF-GA-2013-622527,
Horizon 2020 ERC-Advanced Grants No.~267104 and No.~884719,
Horizon 2020 ERC-Consolidator Grant No.~819127,
Horizon 2020 Marie Sklodowska-Curie Grant Agreement No.~700525 ``NIOBE''
and
No.~101026516,
and
Horizon Europe Marie Sklodowska-Curie Staff Exchange project JENNIFER3 Grant Agreement No.~101183137 (European grants);
L’Institut National de Physique Nucl\'eaire et de Physique des
Particules (IN2P3) du CNRS under Project Identification No.
CNRS-IN2P3-14-PP-033
and L’Agence Nationale de la Recherche (ANR) under Grant No. ANR-23-CE31-
0018 and ANR-25-CE31-1333 (France);
BMFTR, DFG, HGF, MPG, and AvH Foundation (Germany);
Department of Atomic Energy under Project Identification No.~RTI 4002,
Department of Science and Technology,
and
UPES SEED funding programs
No.~UPES/R\&D-SEED-INFRA/17052023/01 and
No.~UPES/R\&D-SOE/20062022/06 (India);
Israel Science Foundation Grant No.~2476/17,
U.S.-Israel Binational Science Foundation Grant No.~2016113, and
Israel Ministry of Science Grant No.~3-16543;
Istituto Nazionale di Fisica Nucleare and the Research Grants BELLE2,
and
the ICSC – Centro Nazionale di Ricerca in High Performance Computing, Big Data and Quantum Computing, funded by European Union – NextGenerationEU;
Japan Society for the Promotion of Science, Grant-in-Aid for Scientific Research Grants
No.~16H03968,
No.~16H03993,
No.~16H06492,
No.~16K05323,
No.~17H01133,
No.~17H05405,
No.~18K03621,
No.~18H03710,
No.~18H05226,
No.~19H00682, 
No.~20H05850,
No.~20H05858,
No.~22H00144,
No.~22K14056,
No.~22K21347,
No.~23H05433,
No.~26220706,
and
No.~26400255,
and
the Ministry of Education, Culture, Sports, Science, and Technology (MEXT) of Japan;  
National Research Foundation (NRF) of Korea Grants 
No.~2021R1-A6A1A-03043957,
No.~2021R1-F1A-1064008, 
No.~2022R1-A2C-1003993,
No.~2022R1-A2C-1092335,
No.~RS-2016-NR017151,
No.~RS-2018-NR031074,
No.~RS-2021-NR060129,
No.~RS-2023-00208693,
No.~RS-2024-00354342
and
No.~RS-2025-02219521,
Radiation Science Research Institute,
Foreign Large-Size Research Facility Application Supporting project,
the Global Science Experimental Data Hub Center, the Korea Institute of Science and
Technology Information (K25L2M2C3 ) 
and
KREONET/GLORIAD;
Universiti Malaya RU grant, Akademi Sains Malaysia, and Ministry of Education Malaysia;
Frontiers of Science Program Contracts
No.~FOINS-296,
No.~CB-221329,
No.~CB-236394,
No.~CB-254409,
and
No.~CB-180023, and SEP-CINVESTAV Research Grant No.~237 (Mexico);
the Polish Ministry of Science and Higher Education and the National Science Center;
the Ministry of Science and Higher Education of the Russian Federation
and
the HSE University Basic Research Program, Moscow;
University of Tabuk Research Grants
No.~S-0256-1438 and No.~S-0280-1439 (Saudi Arabia), and
Researchers Supporting Project number (RSPD2025R873), King Saud University, Riyadh,
Saudi Arabia;
Slovenian Research Agency and Research Grants
No.~J1-50010
and
No.~P1-0135;
Ikerbasque, Basque Foundation for Science,
State Agency for Research of the Spanish Ministry of Science and Innovation through Grant No. PID2022-136510NB-C33, Spain,
Agencia Estatal de Investigacion, Spain
Grant No.~RYC2020-029875-I
and
Generalitat Valenciana, Spain
Grant No.~CIDEGENT/2018/020;
The Knut and Alice Wallenberg Foundation (Sweden), Contracts No.~2021.0174, No.~2021.0299, and No.~2023.0315;
National Science and Technology Council,
and
Ministry of Education (Taiwan);
Thailand Center of Excellence in Physics;
TUBITAK ULAKBIM (Turkey);
National Research Foundation of Ukraine, Project No.~2020.02/0257,
and
Ministry of Education and Science of Ukraine;
the U.S. National Science Foundation and Research Grants
No.~PHY-1913789 
and
No.~PHY-2111604, 
and the U.S. Department of Energy and Research Awards
No.~DE-AC06-76RLO1830, 
No.~DE-SC0007983, 
No.~DE-SC0009824, 
No.~DE-SC0009973, 
No.~DE-SC0010007, 
No.~DE-SC0010073, 
No.~DE-SC0010118, 
No.~DE-SC0010504, 
No.~DE-SC0011784, 
No.~DE-SC0012704, 
No.~DE-SC0019230, 
No.~DE-SC0021616, 
No.~DE-SC0022350, 
No.~DE-SC0023470; 
and
the Vietnam Academy of Science and Technology (VAST) under Grants
No.~NVCC.05.02/25-25
and
No.~DL0000.05/26-27.

These acknowledgements are not to be interpreted as an endorsement of any statement made
by any of our institutes, funding agencies, governments, or their representatives.

We thank the SuperKEKB team for delivering high-luminosity collisions;
the KEK cryogenics group for the efficient operation of the detector solenoid magnet and IBBelle on site;
the KEK Computer Research Center for on-site computing support; the NII for SINET6 network support;
and the raw-data centers hosted by BNL, DESY, GridKa, IN2P3, INFN, 
and the University of Victoria.

\bibliographystyle{JHEP}
\bibliography{biblio.bib}

\end{document}